\documentclass[12pt,preprint]{aastex}

\usepackage{natbib}
\usepackage{amsmath}
\def\ls{{_<\atop^{\sim}}}
\def\gs{{_>\atop^{\sim}}}

\begin{document}

\title{Studying the WHIM Content of the Galaxy Large--Scale Structures along the Line of Sight to H~2356-309}
\shorttitle{Studying the WHIM on the line of sight to H~2356-309}
\shortauthors{Zappacosta et al.}
\author{L. Zappacosta\altaffilmark{1,2}, F. Nicastro\altaffilmark{1,3,4}, R. Maiolino\altaffilmark{3}, G. Tagliaferri\altaffilmark{5},
  D.A. Buote\altaffilmark{6}, T. Fang\altaffilmark{6},
  P.J. Humphrey\altaffilmark{6}, F. Gastaldello\altaffilmark{7}}
\affil{Harvard-Smithsonian Center for Astrophysics, Cambridge, MA,
  02138, USA}
\altaffiltext{1}{Harvard-Smithsonian Center for Astrophysics, 60
  Garden Street, Cambridge, MA 02138; lzappacosta@cfa.harvard.edu}
\altaffiltext{2}{INAF - Osservatorio Astronomico di Trieste, via
  G.B. Tiepolo 11, I-34143, Trieste, Italy.}
\altaffiltext{3}{INAF - Osservatorio Astronomico di Roma, via di Frascati 33, 00040 Monte Porzio Catone, Italy}
\altaffiltext{4}{IESL, FoundationforResearchandTechnology, 71110,
  Heraklion, Crete (Greece)}
\altaffiltext{5}{INAF - Osservatorio Astronomico di Brera, via Bianchi
  46, 23807 Merate (LC), Italy}
\altaffiltext{6}{Department of Physics and Astronomy, 4129 Frederick
  Reines Hall, University of California, Irvine, CA 92697}
\altaffiltext{7}{INAF - IASF, via Bassanini 15, I-20133 Milano, Italy;
  Occhialini Fellow}
\begin{abstract}
We make use of a 500~ks Chandra HRC-S/LETG spectrum of the blazar H~2356-309, combined with a lower S/N (100 ks) pilot 
LETG spectrum of the same target, to search for the presence of warm-hot absorbing gas associated with two Large--Scale Structures (LSSs) crossed by this sightline, and to constrain its physical state and geometry. 
Strong (logN$_{OVII} \ge 10^{16}$ cm$^{-2}$) OVII K$\alpha$ absorption associated with a third LSS crossed by this line of sight 
(the Sculptor Wall, SW), at $z=0.03$, has already been detected in a
previous work.  
Here we focus on two additional prominent filamentary LSSs along the same line of sight, at z=0.062 (the Pisces-Cetus Supercluster, PCS) and at z=0.128 (the ``Farther Sculptor Wall'', FSW). 

The combined LETG spectrum has a S/N of $\sim 11.6-12.6$  per resolution element in the $20-25$ \AA, and an average  
3$\sigma$ sensitivity to intervening OVII K$\alpha$ absorption line equivalent widths of EW$_{OVII} \gs 14$ m\AA\ in the available 
redshift range ($z<0.165$). 
No statistically significant (i.e. $\ge 3\sigma$) individual
absorption is detected from any of the strong He- or H-like transitions 
of C, O and Ne (the most abundant metals in gas with solar-like composition) at the redshifts of the PCS and FSW structures, 
and down to the above EW thresholds. However we are still able to constrain the physical and geometrical 
parameters of the putative absorbing gas associated with these
structures, by performing joint spectral fit of various marginal
detections and upper limits of the 
strongest expected lines with our self-consistent hybrid ionization WHIM spectral model. 

At the redshift of the PCS we identify a warm phase with $\rm \log{T}
=5.35_{-0.13}^{+0.07} ~K$ and $\rm \log{N_H} = (19.1 \pm 0.2)
~cm^{-2}$ possibly coexisting with a much hotter and statistically less significant
phase with $\rm \log{T} = 6.9^{+0.1}_{-0.8} ~K$ and $\rm \log{N_H} =
20.1^{+0.3}_{-1.7} ~cm^{-2}$ ($1\sigma$ errors).
These two separate physical phases are identified through, and mainly constrained by, CV K$\alpha$ (warm phase) and OVIII K$\alpha$ (hot phase) absorption, with single line significances of $1.5\,\sigma$ each. 

For the second LSS, at $z\simeq 0.128$, only one hot component is hinted in the data, through OVIII K$\alpha$ 
($1.6\,\sigma$) and NeIX K$\alpha$ ($1.2\,\sigma$). For this system, we estimate $\rm \log{T} =6.6_{-0.2}^{+0.1} ~K$ and 
$\rm \log{N_H} = 19.8_{-0.8}^{+0.4} ~cm^{-2}$. 

Our column density and temperature constraints on the warm-hot gaseous
content of these two LSSs, combined with the measurements obtained for
the hot gas permeating the SW, allow us to estimate the cumulative
number density per unit redshifts of OVII WHIM absorbers at 3
different equivalent width thresholds of $0.4 \rm m\AA$, $7 \rm m\AA$
and $25.8 \rm m\AA$. This is consistent with expectations only at the
very low end of EW thresholds, but exceed predictions at 7m\AA\ and
25.8 m\AA (by more than $2 \sigma$).
We also estimate the cosmological mass density of the WHIM based on the 4 absorbers we tentatively detect along this line of sight, 
obtaining $\Omega_b^{WHIM} = (0.021^{+0.031}_{-0.018})
(Z/Z_{\odot})^{-1}$, consistent with the cosmological mass
density of the intergalactic 'missing baryons' only if we assume
high metallicities ($Z\sim Z_{\odot}$). 
\end{abstract}


\section{Introduction}
Cosmological hydrodynamical simulations predict the gradual formation  
of a local ($\rm{z}<1$) web of low density 
($n_b=10^{-6}-10^{-5}\,cm^{-3}$), warm-hot ($\rm{T}=10^5-10^7\, 
\rm{K}$) intergalactic gas, connecting
virialized halos (i.e. galaxies, galaxy groups and clusters of galaxies), permeating the large--scale structures  
(LSS) of which these systems are constituents,
and ultimately providing the necessary fuel for their continuous growth  
\citep{cen95,cen,dave,cen06}.
This low-redshift intergalactic gas is largely the same primordial gas  
present at redshift higher than $\sim 2$ in a
cool photo-ionized phase (the so called 'Lyman-$\alpha$ Forest)', but  
in a much hotter and metal-enriched phase, because of
efficient shock-heating during the continuous process of LSS assembly  
and growth in a non-linear Universe, and of strong
feedback with the same structures for which it provides building blocks.
Due to its high temperatures this IGM phase has been dubbed Warm-Hot  
Intergalactic
Medium (WHIM).

Electrons and baryons in the WHIM are shock-heated during their infall  
in the dark matter LSS potential well,
and settle in filamentary/sheet-like structures surrounding LSSs. Such  
matter is predicted to account for a sizable
fraction ($\sim50\%$) of all the baryons in the local ($z<1$)  
universe, and it is thus considered the best candidate
to host the baryons seen at high redshift and missing from the low  
redshift census \citep{fukugita98,fukugita}.

Given its high temperature the WHIM can only emit or absorb in the Far  
UV and soft X-rays, mainly through Li- through H-like
metal transitions and bremsstrahlung continuum emission.
However, at WHIM densities both line and continuum emission are highly  
depressed (due to the dependency of these
mechanisms on the square power of the emitters volume densities), and on  
average well below the sensitivity of current
instrumentations.
Nonetheless, statistical techniques based on cross correlation of  
regions with excess diffuse X-ray emission in ROSAT,
{\em Chandra} and XMM-{\em Newton} deep exposures, with large--scale  
galaxy distribution, have probably already allowed
the marginal detection of the density peak of the WHIM distribution  
\citep[e.g.][]{scharf,zappacosta,zappacosta2,werner}.

A far more promising way to detect the WHIM is through resonant  
absorption by highly ionized metals.
Intervening WHIM filaments should imprint a 'forest', i.e. the so  
called X-ray Forest \citep{hellsten,perna}, by analogy
with the HI Ly~$\alpha$ Forest copiously seen at $z\gtrsim 2$, of metal  
absorption lines onto the spectra of bright background sources,
whose strength depends only linearly on the absorber density, and is  
therefore less suppressed than the corresponding
emission. Predicted equivalent widths (EWs) from the most abundant  
ions in WHIM, range from 1 m\AA\ to $\lesssim 20$ m\AA\
for the densest filaments. 
The detection of even the strongest of such absorption lines (probing  
only the very high density tail of the WHIM distribution: overdensity $ 
\delta = n_b/<n_b> \gtrsim 300$, compared to the average density of the  
Universe: $<n_b>= 2 \times 10^{-7}  (1+z)^3 (\Omega_bh^2/0.02)$, where
$\Omega_b$ is the baryonic density parameter and $h$ is the Hubble
constant in units of 100~km/s/Mpc) is  
extremely challenging with the limited sensitivity ($A_{eff}<60cm^2$)  
and resolution (R=$E/\Delta E=400-800$) of the current {\em Chandra}  
and XMM-{\em Newton} high resolution X-ray gratings.
A $\ge 3\sigma$ detection of an EW=20 m\AA\ absorption line, requires  
spectra of the background targets with S/N$\gtrsim 8$ per 50 m\AA\  
resolution element (i.e. $\gtrsim 70$ net counts per bin). These can only  
be obtained, for the brightest ($\sim 10^{-11} erg\,s^{-1}\,cm^{-2}$)  
soft X-ray  ($\rm0.5-2\,keV$) targets in the extragalactic sky  
(preferably blazars, because of their intrinsically featureless  
spectra), with $\gtrsim 0.5$ Ms  exposures, while in quiescence, or 
$\gtrsim 100$~ks exposures in outburst.

Such dense WHIM filaments are rare. According to hydrodynamical  
simulations, at $z\simeq 0$ one expects $\le 0.05$ WHIM filaments with
EW(OVII$_{K\alpha}$)$\ge 20$ m\AA, per unit redshift. The probability  
of having one of such filaments along a random line of sight
up to $z=0.3$, is therefore only $1.5$\%   
\citep{gehrels}, and only few $z\ge 0.3$ targets with quiescent
F(0.5-2 keV)$\ge 10^{-11}$ erg s$^{-1}$ cm$^{-2}$ are available (e.g.  
Conciatore et al., 2009, priv. comm.).
The number density of OVII WHIM absorbers per unit redshift increases  
by almost two orders of magnitude by lowering the EW(OVII$_{K\alpha})$  
detection threshold down to $\ge 2$ m\AA\, so dramatically increasing  
the probability  of randomly detecting one of such filaments to
P($z=0.3$)=97.5\%.
This observational strategy has the advantage to probe, in theory, the  
bulk of the WHIM mass distribution, but requires incredibly high S/N  
spectra ($\ge 75$ per resolution element with the {\em Chandra} LETG,  
and $\ge 180$ per resolution element with the XMM-{\em Newton} RGS,  
for a $\ge 3\sigma$ detection), obtainable only with several Ms  
exposures on the brightest possible $z\ge 0.3$ targets in their  
quiescent states.
Moreover, even when such high S/N spectra are obtained \citep[for example by  
observing background target during exceptionally high outburst:][]{nicastro05apj,nicastro05} the clear assessment of the statistical  
significance of such weak 2-3 m\AA\ absorption lines is hampered by 
our limited knowledge of the instrument systematics, which is  
comparable to the statistical uncertainties on the lines EWs
\citep{kaastra,rasmussen07,nicastro08}

\begin{figure*}[!]
   \begin{center}
     \includegraphics[scale=0.7]{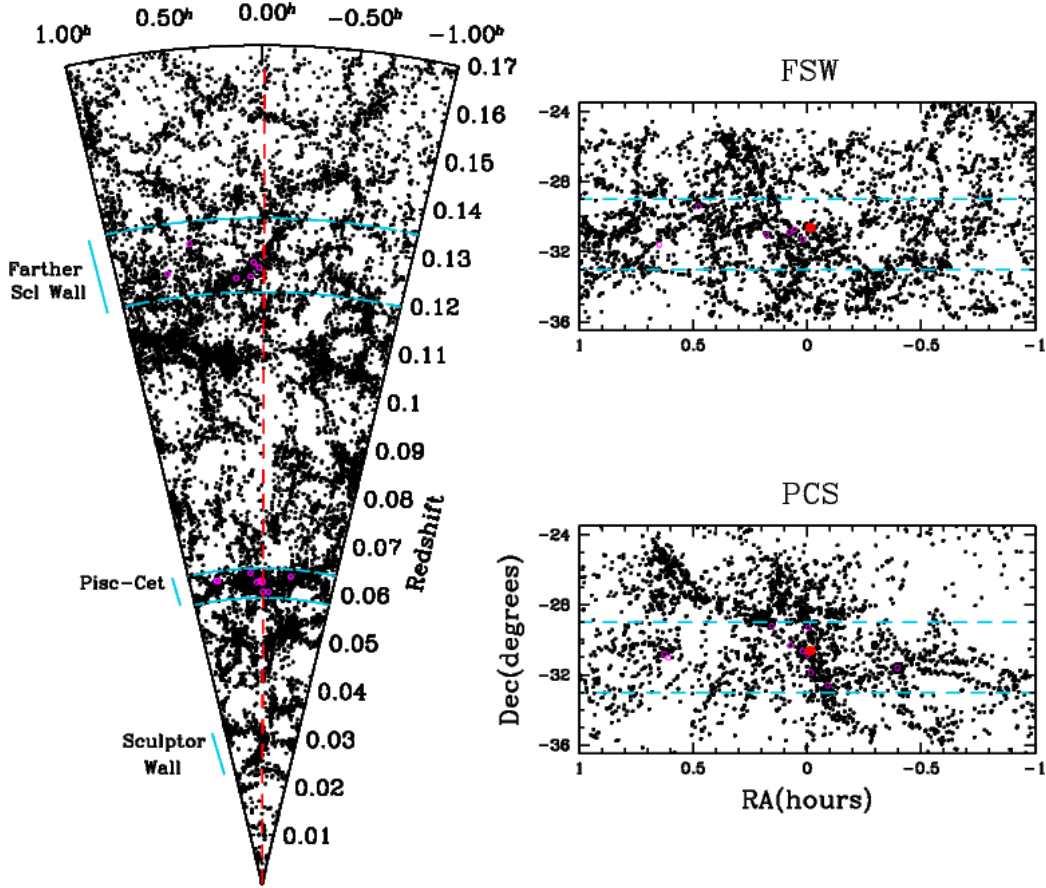}
   \end{center}
\caption{Sky map and wedge diagram of the region of the Sculptor Wall  
where the blazar H~2356-309 is located. The upper sky map refer to
the FSW and the lower to the PCS. Galaxies (black point) clusters and  
groups of galaxies (magenta circles) in the wall are taken from NED.
The wedge diagram show galaxies inside the dashed blue box drawn in  
the sky map and report the sightline to the blazar as red dashed line.  
The
  galaxy and cluster catalogs contain objects belonging to different  
parent catalogs. Hence we point out that some holes in the projected  
galaxy distribution are artificially caused by this non homogeneity  
like the hole at north-east of the blazar position visible at $\rm  
RA=0.1 hours$ and $\rm DEC=-29.5 deg$ in the PCS sky map.}
   \label{LSS}
\end{figure*}

An alternative observational strategy is to select lines of sight  
where the probability to cross dense WHIM filament is enhanced \citep[e.g.][]{zapproposal,zapproposal2}.
WHIM gas density and metallicity is predicted \citep[and possibly in the UV  
observed][]{stocke} to correlate
with LSS galaxy overdensity \citep{viel}. Thus, chances of intercepting a dense  
WHIM filament may be improved by carefully selecting bright
sources in the background of extreme filamentary LSS concentrations  
(Fig.~\ref{LSS}).
This observational strategy has been successfully exploited recently  
in \citet{buote09}, who detected a strong (EW$\sim30 m\AA$)
absorption line in two XMM-{\em Newton} RGS and {\em Chandra}  
LETG spectra of the blazar H~2356-309 (z=0.165), and tentatively identified it
with OVII K$\alpha$ at a redshift consistent with that spanned by the  
intervening Sculptor Wall \citep[SW;][]{daCosta}.

Here, and in our companion paper \citep{fang10}, we report on the  
successful extension of this observational program.
H~2356-309 has been recently re-observed with the {\em Chandra} LETG  
for 500 ks, as part of an approved cycle 10 
GO observational program.
The main goal of this deep observation was to confirm, at higher  
significance, the SW OVII detection \citep{fang10}.
Secondary objectives, were to confirm the presence of other (lower
significance) lines from the same absorber (Buote et al., 2010 in prep.). 
Here instead we focus on constraining the physical parameters of the  
putative WHIM gas content of two additional
galaxy LSSs present along this line of sight, at $z=0.062$ (the
Pisces-Cetus Supercluster; PCS)  
and $z=0.128$ (a farther wall which we will call Farther Sculptor
Wall; FSW).
In this paper we use all the existing {\em Chandra} LETG data of  
H~2356-309, to characterize the physical properties of the WHIM
permeating these additional structures, and conclude by estimating the  
contribution of such dense gaseous component of LSSs, to the
WHIM cosmological mass density.
In \S\ref{sightline} we describe the richness of LSSs along the line  
of sight to H~2356-309.
In \S\ref{datareduction} and \S\ref{analysis} we present the data and  
describe their reduction and analysis.
\S\ref{discussion} is devoted to a critical discussion of our findings.
In \S\ref{conclusions} we summarize our conclusions.
Throughout the paper we adopt a $\Lambda$-CDM cosmology, with $h=0.71$, $\Omega_M=0.27$, $\Omega_{\Lambda}=0.73$.

\section{The LSS Richness of the Line of Sight to  
H~2356-309}\label{sightline}
Figure~\ref{LSS} shows the wedge diagram of the line of sight to  
H~2356-309 in the declination range
$-33 < \delta < -29$. Galaxies (black point), clusters and groups of  
galaxies (magenta circles) shown in the diagrams of Fig.~\ref{LSS},
are extracted from a number of different and non-homogeneous catalogs  
and galaxy surveys (including the 2dF Galaxy Redshift Survey - 2dFGRS,
\citealt{colless}; and 6dF Galaxy Redshift Survey - 6dFGRS, \citealt{6dF}), and are the result of a general query to the  
Nasa/Ipac Extragalactic Database (NED)\footnote{The NASA/IPAC Extragalactic Database (NED) is operated by the Jet Propulsion Laboratory, California Institute of Technology, under contract with the National Aeronautics and Space Administration.}. As such, these diagrams
do not represent complete flux-limited sample of the actual galaxy  
distribution along this line of sight.

Several strong LSS concentrations are clearly visible, and cross the  
line of sight to H~2356-309 at, at least,
three different average redshifts: $<z_1>=0.03$ (the SW),  
$<z_2>=0.062$ (the PCS), and $<z_3>=0.128$ (the FSW).
Both the PCS and the FSW LSSs are significantly larger than the SW,  
and are delimited in the wedge diagram of Fig.~\ref{LSS}
by cyan dashed arcs. The two 2D sky map projections on the right of  
the wedge diagram of Fig.~\ref{LSS}, show the RA versus DEC
extent of these two LSSs.

The PCS \citep[][]{burns,tully} is one of the richest nearby (z$< 
$0.1) superclusters. It is clearly visible in the SDSS and 2dF
redshift surveys as a remarkable filament of galaxies \citep{porter}.
The structure intercepted by the line of sight to H~2356-309 is a long  
filament of galaxies located on the plane of
the sky  at z=0.06-0.063, within $<1Mpc$ from the projected blazar  
position.

The FSW is a conspicuous wall of galaxies, which originates from  
the Sculptor Supercluster at z=0.11, and stretches out
to redshift $z\sim0.135$ crossing the blazar sightline at $\Delta  
z=0.127-0.129$.

\section{Observations and data reduction}\label{datareduction}
The blazar H~2356-309  has been observed twice with the {\em Chandra}  
LETG, as part of two different observational programs.
A first 100 ks LETG observation was performed in October 2007 
and has been already analyzed in \citet{buote09}. 
A second, deeper LETG observation was carried out over the 
September--December 2008 period, through
ten different pointings with exposures ranging from 15 ks to 100 ks,  
for a total of 496.4ks.
The aim of this observation was to secure, with a conservative flux of
$1.0\times10^{-11} \rm erg s^{-1} cm^{-2}$ (0.5-2.0~keV) a $5\sigma$
significant detection of an absorber with a column density of at least
$9\times10^{15} cm^{-2}$ (this is the 90\% lower limit found by
\citealt{buote09}) by means of a long 0.5~Ms non-ToO observation. 
Table~\ref{log} shows the log of the observations.

We mention that H~2356-309 has also been observed with XMM for 130~ks
and this observation has been analyzed in \citet{buote09}. We will not
use this observation in our analysis since: 1) it adds only 70 counts in
0.06$\AA$ (the XMM-RGS FWHM) which are $\sim1/4$ the net counts all the
Chandra data provide in 0.05$\AA$; 2) does not reach the wavelengths
of the CV transition which, as we will see in \S\ref{globanalysis},
characterizes the most important intervening system; 3) makes the
analysis in \S\ref{globanalysis} very complicated due to the presence
of several instrumental features very close to the lines we are
investigating. 

\begin{table*}
  \caption{Log of the Chandra observations of H~2356-309.}
  \begin{center}
    \begin{tabular}[t]{|rrrr|}
\hline
\hline
ObsID &  Instrument  & Exposure & Date of observation\\
\hline 
8120  &  HRC-S/LETG  &  96.49  & 2007-10-11 \\
10497 &  HRC-S/LETG  &  53.93  &  2008-09-19 \\                      
10498 &  HRC-S/LETG  &  77.88  &  2008-09-22 \\                      
10499 &  HRC-S/LETG  &  58.69  &  2008-09-29 \\                      
10500 &  HRC-S/LETG  &  16.15  &  2008-12-25 \\                      
10577 &  HRC-S/LETG  &  81.72  &  2008-09-17 \\                      
10761 &  HRC-S/LETG  &  42.15  &  2008-09-27 \\                      
10762 &  HRC-S/LETG  &  35.17  &  2008-09-25 \\                      
10764 &  HRC-S/LETG  &  100.4  &  2008-10-10 \\                      
10840 &  HRC-S/LETG  &  15.14  &  2008-12-23 \\                      
10841 &  HRC-S/LETG  &  15.15  &  2008-12-28 \\                      
\hline
\hline
   \end{tabular}
  \end{center}
\label{log}
\end{table*}

\begin{figure*}[!]
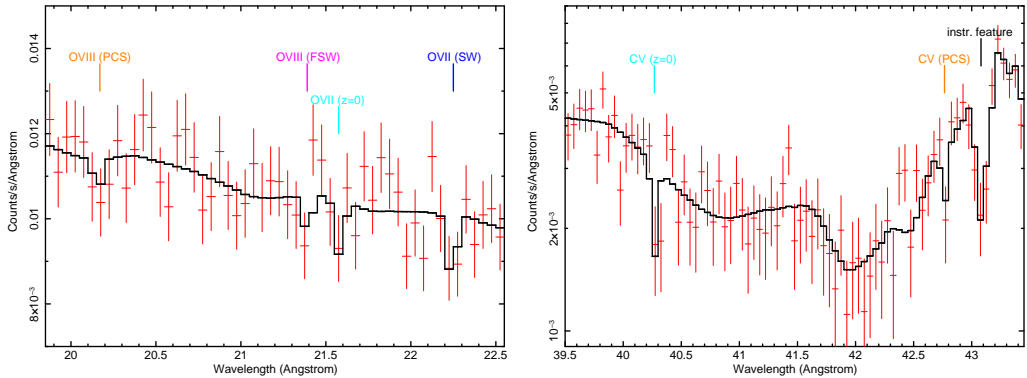

  \begin{center}
   \includegraphics[width=0.3\textwidth, angle=270]{f2a.ps}
    \includegraphics[width=0.3\textwidth, angle=270]{f2b.ps}
   \caption{Two portions of the spectral regions considered in
      \S\ref{singleline}. The left panel shows the Chandra spectrum 
      around the rest frame OVII line region while the right panel
      shows the CV region. Each of the possibly identified lines
      (modelled in the spectrum with a Gaussian line profile) is
      labelled with colors indicating different absorbing systems. The
      fitted redshift of each line is showed in 
      Table~\ref{lines}}
    \label{specsingleline}
  \end{center}
\end{figure*}

\subsection{Chandra reduction}
Each Chandra observation was reduced with the latest version of the {\em Chandra Interactive Analysis of Observation} software (CIAO v. 4.1.2, 
CALDB v. 4.1.2), following the standard processing procedures outlined in
the HRC-S/LETG Grating analysis thread\footnote{http://asc.harvard.edu/ciao/guides/gspec\_hrcsletg.html}, and 
applying a new (and still not standard) filtering procedure\footnote{http://asc.harvard.edu/contrib/letg/GainFilter/} on the level 1 event files. 
This allowed us to greatly reduce the number of background counts, compared to the standard 
pipeline procedure, while losing only a negligible percentage of source counts, and so 
greatly increasing the S/N of the background subtracted source spectrum. 

For each observation, we produced background light curves and inspected them visually to filter out 
periods of background flares. Only few observations were affected by 
short and mild flaring periods and their screening removed a total of just 30 ks, leaving a cleaned total 562 ks exposure.  

To maximize the throughput, we used the {\em Chandra} LETG in combination with the {\em Chandra} High Resolution Camera for Spectroscopy (HRC-S) dispersion detector. This has virtually no spectral resolution, which prevents the separation of the different spectral orders dispersed by the LETG. We extracted 'all-order' negative and positive source and background spectra with the CIAO tool {\em tg\_extract}. 
Due to impossibility of separating HRC-LETG spectral orders, the spectral modeling of these spectra can only be performed by pre-folding the fitting models with the sum of the convolution products of the redistribution matrices (RMFs) and ancillary responses files (ARFs) belonging to the first N positive and negative orders, respectively. Here N is a number that depends on the intensity of the source, its spectral energy distribution, and the possible presence of strong spectral absorption or emission features, and has to be carefully chosen on an observation by observation base. 
For our LETG observations of H~2356-309, we verified\footnote{http://cxc.harvard.edu/ciao/threads/hrcsletg\_orders/} that in no case high-energy photons
from orders higher than 6 contaminate the low energy spectrum by more
than 3\% (0.06\% at the OVII wavelength). We decided to conservatively build our final ``all-order'' response matrix by adding up positive and negative orders up to N=10.  
For each order, we used the ftools task MARFRMF\footnote{http://heasarc.gsfc.nasa.gov/lheasoft/ftools/caldb/marfrmf.html} to 'multiply' the RMF by its corresponding ARF. This produced 10 positive and 10 
negative normalized RMFs, which we then co-added with the ftools task ADDRMF\footnote{http://heasarc.gsfc.nasa.gov/lheasoft/ftools/caldb/addrmf.html}. 

Finally, to maximize the S/N of our spectra, for each observation we co-added negative and positive order source and background spectra 
(add\_grating\_orders task) and response files (ADDRMF).

Because of our co-adding procedure, particular care must be devoted to
account for the wavelength calibration inaccuracy. An updated version of the
degap polynomial coefficients based on empirical wavelength corrections from
multiple HRC-S/LETG observations of Capella has been release since
CALDB~3.2\footnote{http://cxc.harvard.edu/cal/Letg/Hrc\_disp/degap.html}.
This update allow the correction of the non-linearities in the HRC-S
dispersion relation improving the uncertainties across the detector
from $0.014 \rm\AA$ to $0.010 \rm\AA$ (RMS deviation). This
uncertainty is lower at shorter wavelengths and higher at longer
wavelengths. Since we have coadded positive and negative orders we
should consider a larger uncertainty. The propagation of the RMS deviations
gives an uncertainty of $0.014 \rm\AA$.  However to be conservative we
adopt $20\rm m\AA$ of uncertainty since the updated wavelength
corrections have been applied to the rest frame position of the
strongest soft X-ray metal electronic transitions and the
interpolation of the correction may not be strictly valid to the position of
blue- and red-shifted lines.

\section{Spectral Analysis}\label{analysis}
We performed all our spectral analysis with the fitting package {\em
  Sherpa} of CIAO (v. 4.1.2). The statistics we used for our fits is
the data weighted $\chi^2$ with the Gehrels variance function 
\citep{gehrels}, the default in {\em Sherpa}. 
We first checked for variability of the broad band 10-50\AA\ continuum
(flux and spectral shape) of H~2356-309, between the 11 {\em Chandra}
observations. We grouped each spectrum at a minimum of 20 counts per
bin, and modeled each data-set independently with a power law
attenuated by the sightline Galactic column of neutral gas ($N_H =
1.44 \times 10^{20}$ cm$^{-2}$; \citealt{karlberla}).  We found that
both the source spectral shape and flux varied only moderately (12\%
and 20\%, respectively) between the ten 2008 observations: the 0.3-1.0
keV flux ranges within the $1.25-1.5\times10^{-11} erg s^{-1} cm^{-2}$
interval, while the power law best fitting spectral indices vary
between $\Gamma=1.96-2.20$.  The 0.3-1 keV source flux during the 2007
observation, is 50-80\% lower than during the 2009 observations, but
the best fitting power law spectral index of $\Gamma = 2.19\pm0.06$ is
still in the range measured during the 2008 observations.  Since both
the 10-50 \AA\ flux and spectral shape of the target varied only
moderately between the different observations, and because we are
interested in the search of narrow spectral features, which are not
affected by broad-band spectral-shape variability, we decided to
co-add the HRC-S/LETG spectra of all the observations, to increase the
final S/N per resolution element.  The resulting spectrum has $\simeq
290$ net source Counts per 50 m\AA\ Resolution Element (CPRE) and
$\sim120-160$ background CPRE at 22 \AA\ , and
so a S/N$\sim 11..6-12.6$ at 22 \AA. This gives a theoretical $\ge 1\sigma$
sensitivity to absorption lines with EW$\ge 12-13$ m\AA, at 22 \AA.  We
grouped the data to half the nominal FWHM of LETG 
spectra with $25$ m\AA\ per bin for all the subsequent analysis.

\subsection{Search for Intervening Absorption Lines}\label{singleline}
Our main goal is to constrain the warm-hot gaseous content of the two filamentary galaxy structures here identified as PCS and FSW. 
At the expected WHIM temperatures the most intense absorption lines are the K$\alpha$ resonant transitions from He- and H-like 
C, O and Ne. Table~\ref{explines} lists the rest frame wavelengths and oscillator strengths of these transitions. 

\begin{table}[t]
   \caption{Strongest Metal Transitions in Gas at T=$10^5-10^7$ K}   
   \begin{center}
     \begin{tabular}{|lcc|}
\hline
\hline
Ion     &  Wavelength & Oscillator Strength   \\
\hline
CV K$\alpha$     &    40.2678  &    0.648    \\
CVI K$\alpha$    &    33.7360  &    0.416    \\
OVII K$\alpha$   &    21.6019  &    0.696    \\
OVIII K$\alpha$  &    18.9689  &    0.416    \\
NeIX K$\alpha$   &    13.4471  &    0.724    \\
NeX  K$\alpha$   &    12.1339  &    0.416    \\  
\hline       
     \end{tabular}
\label{explines}
   \end{center}
 \end{table}

A crucial condition for the search for unresolved absorption lines in intrinsically featureless spectra, is the accurate subtraction of the local continuum. 
A visual inspection of the broad band, 10-50 \AA, residuals of the co-added LETG spectrum of H~2356-309, after subtracting the best-fitting absorbed power law, clearly showed the presence of 1-3 \AA\ broad systematics, particularly near or at the wavelength of the main instrumental edges of CI and OI. In general, the residuals deviated significantly from zero, in both directions, in several spectral regions. 
This makes the assessment of the actual significance of a possible absorption line at such wavelengths, difficult. 
For each of the two redshifted systems (the PCS and the FSW), we therefore decided to isolate four 3-6 \AA\ broad spectral regions, around the absorption lines of interest (Table 1), at the following rest-frame intervals: 11-14 \AA\ (NeX, NeIX), 18.5-25.0 \AA\ (OVII, OVII), 
31.5-34.5 \AA\ (CVI), 39-42 \AA\ (CV). 
We fitted each of these spectral intervals with continuum models including third-order polynomia. We then inspected visually the residuals to 
look for possible relatively broad 0.5-1 \AA\ deviations (due to
instrumental effects), and, when needed, added to the best-fit polynomium 0.5-1 \AA\ broad 
emission or absorption Gaussian, to improve the modelling of the local continua. We iterated this procedure until we obtained a satisfactory 
description of all the local continua, and  a visual inspection of the residuals revealed no further deviations. 

For each of the transitions listed in Table 2, we then added to the local best-fitting continuum a negative-only Gaussian (i.e. with flux allowed to be only negative in the fit), with wavelength allowed to vary within the redshift intervals of the PCS and the FSW and FWHM frozen to 10 m\AA\ (unresolved in the LETG), and re-fitted the data. 
For each line we estimate a single line significance in standard deviations, as the ratio between its best-fitting flux, and its 1$\sigma$ error 
(computed with the {\em Sherpa} routine {\em projection}\footnote{http://cxc.harvard.edu/sherpa/ahelp/projection.py.html}, by leaving both the continua and Gaussian normalizations free to vary). 
Four and three of the transitions listed in Table 2 were preliminarly,
and tentatively, 
identified at a single line significance $> 1\sigma$, for the PCS (CV,
OVII, OVIII, and NeIX) and the FSW (CVI, OVIII and NeIX),
respectively. For the remaining transitions we list their 3$\sigma$ EW
upper limits. For comparison, and completeness, Table 3 also lists the
result of our fitting procedure for the lines listed in Table 2, at
$z\simeq 0$. 
five of these transitions are detected at $\ge 1\sigma$ at $z\simeq
0$ (CV, OVII, OVIII, NeIX and NeX). What discussed above provides only a preliminary attempt to 
check for the presence of the expected absorption lines at the redshifts of the 
Large--Scale galaxy Structures present along the lines of sight. 
The presence (or absence) of a given line (at the quoted significance level: 
Table~\ref{lines}) is only used as a guidance for the detailed global fitting procedure 
presented in next section, and that that makes use of our self-consistent WHIM 
collisional ionization plus photoionization model.

\begin{table*}[t]
   \caption{Best-Fitting Absorption Line Parameters and Ids for the PCS and the FSW}   
   \begin{center}
     \begin{tabular}{|ccc|cc|}
\hline
\hline
Wavelength & EW &  Significance$^a$ & Identification & Redshift$^b$ \\
(in \AA) & (in m\AA) & (in $\sigma$) & & \\
\hline
\multicolumn{5}{|c|} {Pisces-Cetus Supercluster (PCS)} \\
\hline
$42.770 \pm 0.015$ & $19.6 \pm 13.1$ & 1.5 & CV K$\alpha$  &  $0.0621 \pm 0.0004$ \\
35.76 -- 35.86 & $< 5.5$ & NA & CVI K$\alpha$  &  0.060 -- 0.063 \\
$22.970 \pm 0.015$ & $6.2 \pm 5.5$ & 1.1 & OVII K$\alpha$  & $0.0633 \pm 0.0008$ \\
$20.160 \pm 0.015$ & $7.3 \pm 4.8$ & 1.5 & OVIII K$\alpha$  &  $0.0628 \pm 0.0007$\\
$14.280 \pm 0.015$ & $5.0 \pm 3.3$ & 1.5 & NeIX K$\alpha$  &  $0.062 \pm 0.001$ \\
12.86 -- 12.90 & $< 3.2$ & NA & NeX  K$\alpha$  &  0.060 -- 0.063\\
\hline
\multicolumn{5}{|c|} {Farther Sculptor Wall (FSW)} \\
\hline
45.38 -- 45.46 & $< 6.1$ & NA & CV K$\alpha$  &  0.127 -- 0.129 \\
$38.042 \pm 0.015$ & $10.4 \pm 8.5$ & 1.2 & CVI K$\alpha$  &  $0.1278 \pm 0.0004$ \\
24.32 -- 24.39 & $< 5.8$ & NA & OVII K$\alpha$  & 0.126 -- 0.129  \\
$21.396 \pm 0.015$ & $7.7 \pm 4.9$ & 1.6 & OVIII K$\alpha$  & $0.1279 \pm 0.0008$ \\
$15.161 \pm 0.015$ & $4.0 \pm 3.3$ & 1.2 & NeIX K$\alpha$  &  $0.127 \pm 0.001$ \\
12.46 -- 12.47 & $< 3.1$ & NA & NeX K$\alpha$  &  0.127 -- 0.129 \\
\hline
\multicolumn{5}{|c|} {$z \simeq 0$} \\
\hline
$40.274 \pm 0.015$ & $38.5 \pm 12.3$ & 3.1 & CV K$\alpha$  &  $0.0002 \pm 0.0005$\\
33.72 -- 33.75 & $< 7.1$ & NA & CVI K$\alpha$  &  $\pm 0.0004$ \\
$21.561 \pm 0.015$ & $8.1 \pm 5.0.$ & 1.6 & OVII K$\alpha$  &  $-0.0019 \pm 0.0007$ \\
$18.952 \pm 0.015$ & $4.6 \pm 3.6$ & 1.3 & OVIII K$\alpha$  &  $-0.0009 \pm 0.0008$ \\
$13.449 \pm 0.015$ & $7.6 \pm 3.3$ & 2.3 & NeIX K$\alpha$  &  $\pm 0.001$\\
$12.131 \pm 0.015$ & $7.7 \pm 2.9$ & 2.7 & NeX K$\alpha$  &  $\pm 0.001$ \\
\hline 
     \end{tabular}
\label{lines}
   \end{center}
$^a$ Single Line Significance in Standard Deviations, evaluated as the ratio between the best fitting EW and its 1$\sigma$ error (see text for details). 

$^b$ The error is derived from the systematic 1$\sigma$ wavelength uncertainty of $\pm 15$ m\AA\, due to the non-linearity of the HRC-S LETG dispersion relationship. 
 \end{table*}

Figure ~\ref{specsingleline} shows two spectral portions of the total {\em Chandra} LETG spectrum of H~2356-309, containing the 
three lines tentatively identified at the redshifts of the PCS and the FSW. \\

\subsection{Constraining the Physics of the Putative PCS and FSW Absorbers with a Self -Consistent WHIM Model }\label{globanalysis}
In \S\ref{singleline} we generally searched for the presence of the strongest absorption metal lines expected from highly ionized gas with temperatures in the broad interval $T=10^5-10^7$ K. However, these transitions belong to ions with quite different ionization potential (e.g CV and NeX) and whose relative abundances critically depend on the ionization mechanisms at work, and on the actual gas temperature. 
For example, in collisionally ionized gas, OVII, being H-like and so quite stable, is virtually the only abundant ion of O within a broad interval of temperatures, within logT(K)=5.5-6.2. In the same temperature interval the lower ionization ions CV and CVI are relatively abundant, with CV decreasing monotonically from $\sim 75$\% down to $\sim5$\%, and CVI raising from $\sim 25$\% up to $\sim 50$\% at its logT(K)=6 peak temperature, and then decreasing again down to $\sim 20$\%. On the contrary the higher ionization ions OVIII, NeIX and NeX are only important at the high-temperature extreme of this interval, with relative abundances steeply raising from very low values up to 23\% (OVIII), 96\% (NeIX) and 3\%  (NeX). 
Modeling the spectra of extragalactic sources crossing regions of the Universe with large galaxy overdensities (expected to trace WHIM filaments) with self-consistent ionized absorber models, may therefore provide useful constraints on the ionization state and column density of the putative absorbers embedded in these LSSs, even if the spectral signature of the gas are individually marginally detected (and/or upper limits are obtained) as long as they are modelled jointly. 

Here we make use of an adaptation of the Photoionized Absorber Spectral Engine (PHASE; Krongold et al., 2003) code, for WHIM gas \citep[e.g.][]{nicastro10}. The code includes more than 3000 electronic resonant transitions (including metal inner-shell) from all elements lighter than Ni, and computes, for a given H equivalent column density, temperature and turbulence velocity of the absorber, the Voigt-profile folded opacity of each transition. The ionization balance in the gas is computed by perturbating the equilibrium collisional ionization balance at a given temperature T, with photoionization by the meta-galactic UV and X-ray background at a given redshift (the redshift of the absorber). Such second-order photoionization contribution depends uniquely on the baryon density $n_b$ in the gas (the lower the density the higher the contribution of photoionization), and it starts to be effective at $n_b \ls 10^{-5}$ cm$^{-3}$. 

In our fitting procedure, we use the same initial methodology adopted in \S\ref{singleline}. 
For each of the two super-structures, the PCS and the FSW, we fit the 4 different narrow-band portions of the continuum where the main transitions lie, independently. For each spectral interval, our fitting model includes the best-fitting continuum determined in \S\ref{singleline} attenuated by 
our hybrid-ionization absorption WHIM model. 
For each spectral interval we leave free to vary in the fit, the continuum normalization and two out of the five parameters of the WHIM model, namely: the equivalent H column density N$_H$ and the temperature T of the gas. Both N$_H$ and T are linked to their same respective values in the 4 independently fitted spectral regions. 
The remaining parameters of our WHIM model are the turbulence velocity $v$ (summed in quadrature to the thermal Doppler parameter of a given transition), the redshift $z$ and the baryon density $n_b$ of the absorbers. The baryon density $n_b$ is highly degenerate with the electron temperature (which is set mainly by collisions in shock-heated WHIM gas), and modifies only slightly the ionization balance of the gas, compared with pure collisional equilibrium at a given temperature.Consequently, $\rm n_b$ can be constrained independently of temperature only with data where several transitions from different ions of the same element are clearly detected. In our fit of the putative absorbers at the redshifts of the PCS and the FSW, we therefore freeze the gas baryon density to a typical WHIM value of n$_b=10^{-5}$ cm$^{-3}$.  
Analogously, the turbulent velocity of the absorber (degenerate with
the ion column density for saturated lines), can only be properly
constrained when the single absorption lines are resolved and their
profile clearly detected at high significance. We freeze this
parameter to $v=100$ km s$^{-1}$, comparable to typical values of
Doppler velocities inferred by hydrodynamical simulations \citep[e.g. ][]{cenfang}. 
Finally, for each of the 4 spectral regions, we first leave the redshift of the absorbers to vary independently over the entire redshift extent of the two super-structures ($z=0.060-0.063$ for the PCS and $z=0.127-0.129$ for the FSW), and then refine the fitting by freezing the redshift of the absorber in the spectral region where the most significant absorption line is detected to its best fitting value, and leaving the redshifts of the other absorbers in the three remaining spectral regions, free to vary between $\pm 20$ m\AA\ from the frozen redshift of the most significant absorption line, to account for the 90\% systematic wavelength uncertainties in the HRC-S LETG. 

Table 4 summarizes the results of our fit. Errors are quoted at a 68\% confidence limit. 
For each parameter listed in table 4, we compute errors by allowing
the continuum normalization to vary within its $1\sigma$ uncertainty 
and allowing the other free WHIM model parameters free to vary except 
in the case of the estimation of $\log N_H$ and $\log T$ errors where 
we fixed the redshifts to their best-fit value.

\begin{table}
  \caption{Redshift and physical parameters resulting from the fits}
  \begin{center}
    \begin{tabular}[t]{|ccc|}
\hline
\hline
$\log T$ & $\log N_H$ & Redshift \\
(in K) & (in ($Z/Z_{\odot})^{-1}$ cm$^{-2}$) & \\
\hline
\multicolumn{3}{c}{{\bf PCS  (z=0.06-0.063: Warm Phase)}}\\
\hline
$5.35_{-0.13}^{+0.07}$  & $19.1 \pm 0.2$ & $0.0623 \pm 0.0005$ \\
\hline 
\multicolumn{3}{c}{{\bf PCS  (z=0.06-0.063: Hot Phase)}}\\
\hline
$6.9_{-0.8}^{+0.1}$  & $20.1_{-1.7}^{+0.3}$ & $0.063 \pm 0.001$ \\
\hline 
\multicolumn{3}{c}{{\bf FSW  (z=0.127-0.129)}}\\
\hline 
$6.6_{-0.2}^{+0.1}$ & $19.8_{-0.8}^{+0.4}$ & $0.126 \pm 0.001$ \\
\hline
    \end{tabular}
\label{resultsproj}
  \end{center}
\end{table}

\subsubsection{The PCS filament}
\begin{figure}[!t]
  \begin{center}
\includegraphics[width=0.45\textwidth, angle=0]{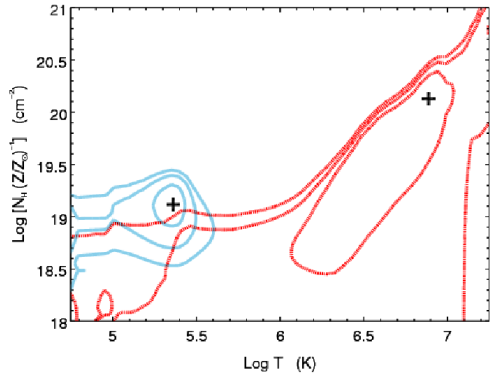}
\caption{68\%, 90\% and 95\%  temperature and equivalent H column density confidence contours for the two putative WHIM phases (warm - solid cyan - and hot - dotted red) permeating the PCS.}
\label{contoursPCS}
  \end{center}
\end{figure}

\begin{figure*}[!t]
  \begin{center}
\includegraphics[width=0.45\textwidth, angle=0]{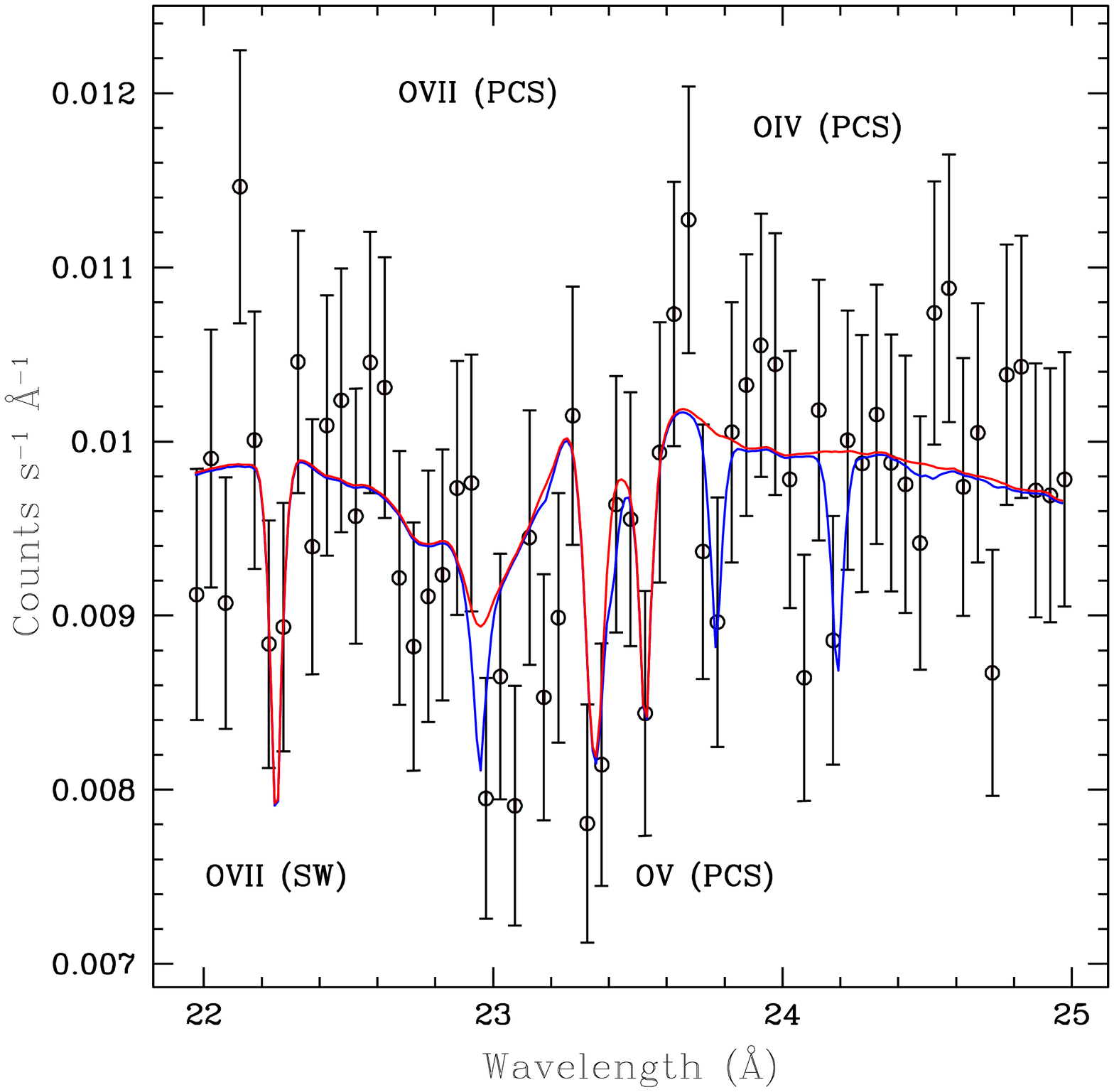}
\includegraphics[width=0.45\textwidth, angle=0]{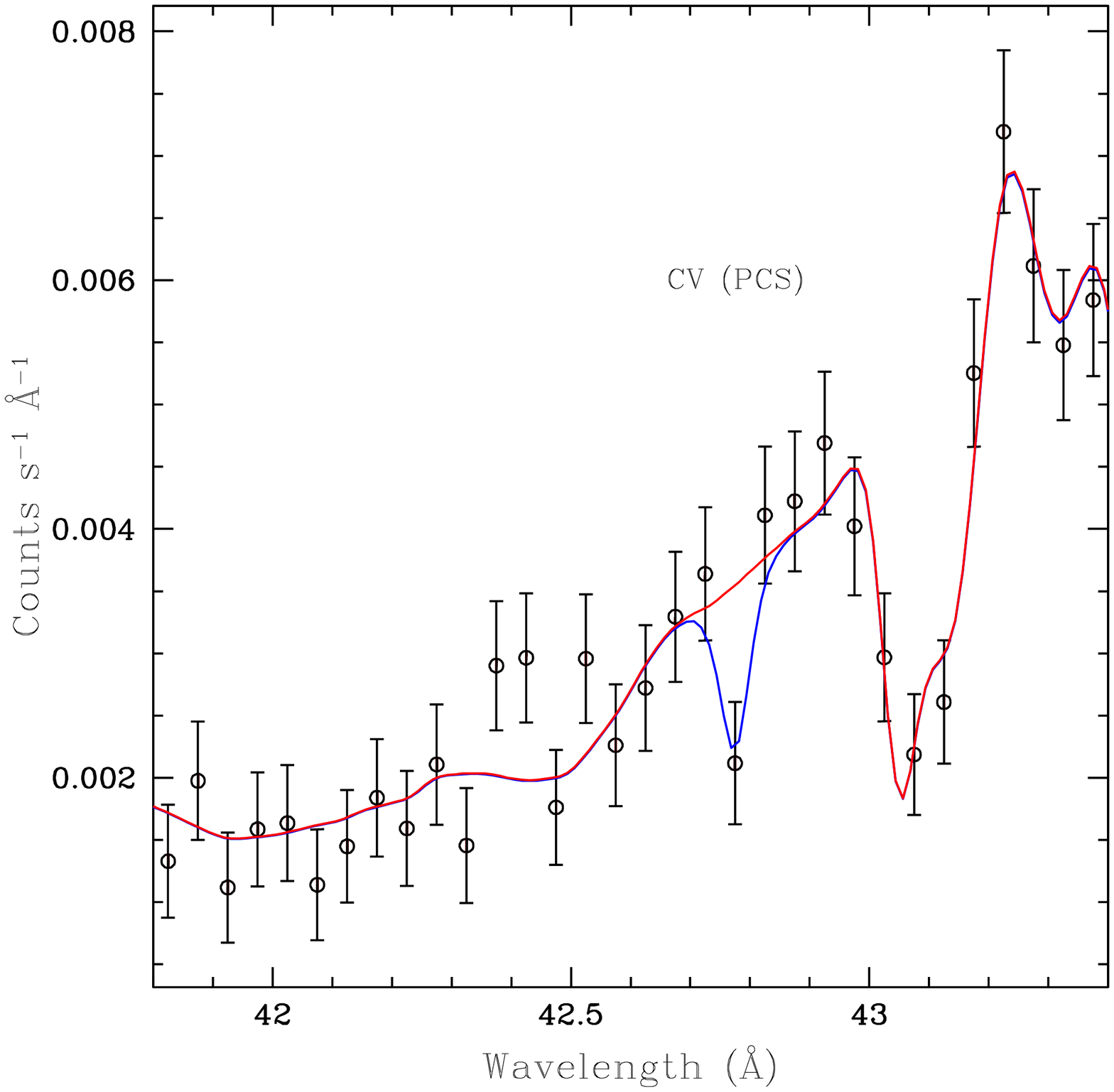}
\caption{Portions of the LETG spectrum of H~2356-309 showing the
  strongest absorption lines from the warm component of the PCS.  The
  best-fitting continuum model is shown in solid thick red, while
  absorption lines from the warm WHIM component are shown in solid
  thin blue. 
The CV K$\alpha$ absorption line at a redshifted wavelength of $\lambda=42.776$ \rm\AA\ is the strongest line of the best-fitting warm-component of the PCS (right panel). 
The other strong lines predicted by the model are the outer shell OVII K$\alpha$ absorption line at $\lambda=22.95\rm\AA$ (superimposed on the OI instrumental edge), and the two inner shell K$\alpha$ transitions from OV at $\lambda=23.74\rm\AA$  ($\lambda=22.35\rm\AA$ rest frame) and OIV at $\lambda=24.17\rm\AA$ ($\lambda=22.75\rm\AA$ rest frame; left panel).  
Also shown in the left panel are the $z\simeq 0$ OVII K$\alpha$ and the atomic and molecular OI K$\alpha$ lines.}.
\label{spectraPHASEwarmPCS}
  \end{center}
\end{figure*}

\begin{figure*}[!t]
  \begin{center}
\includegraphics[width=0.45\textwidth, angle=0]{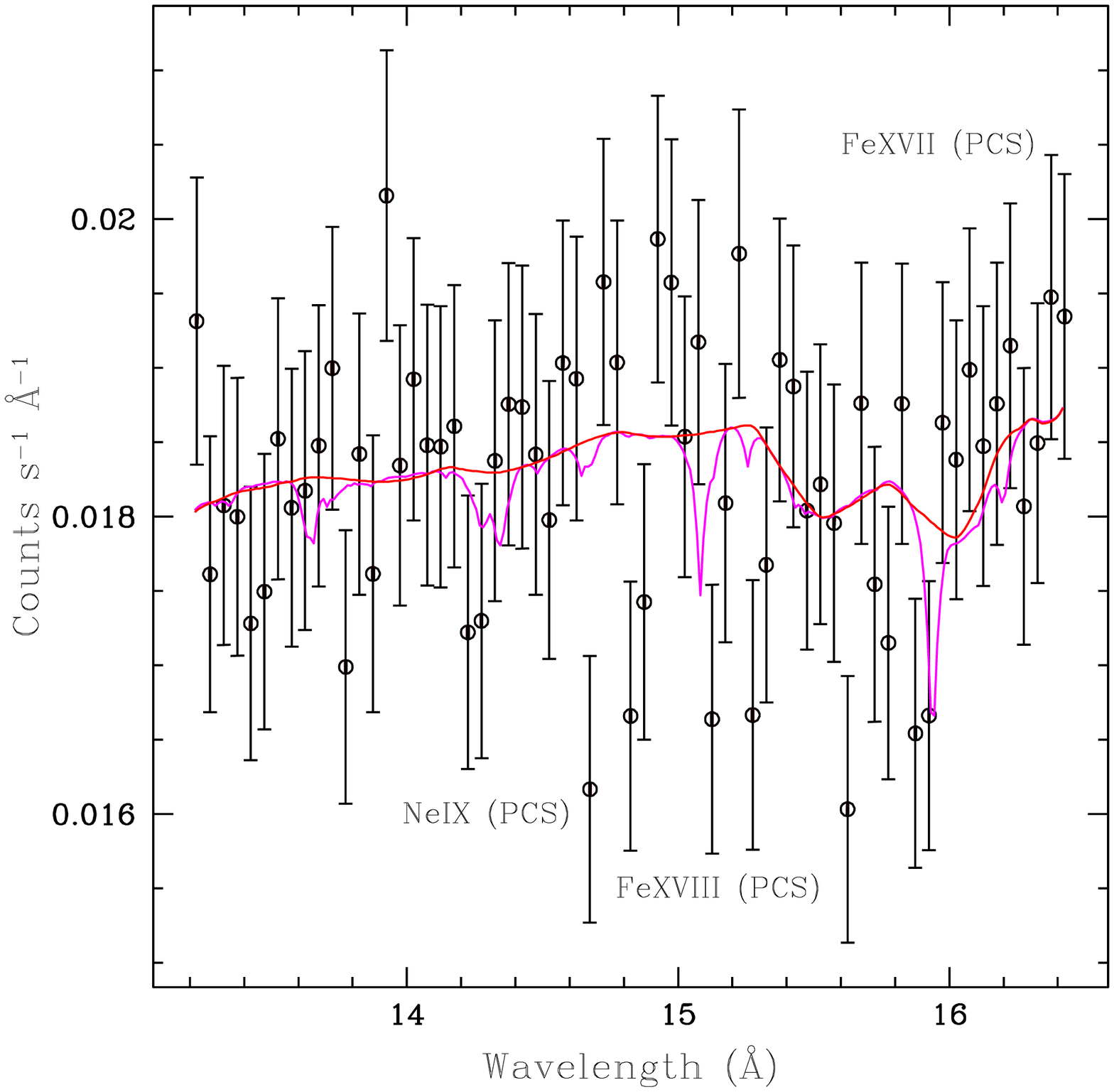}
\includegraphics[width=0.45\textwidth, angle=0]{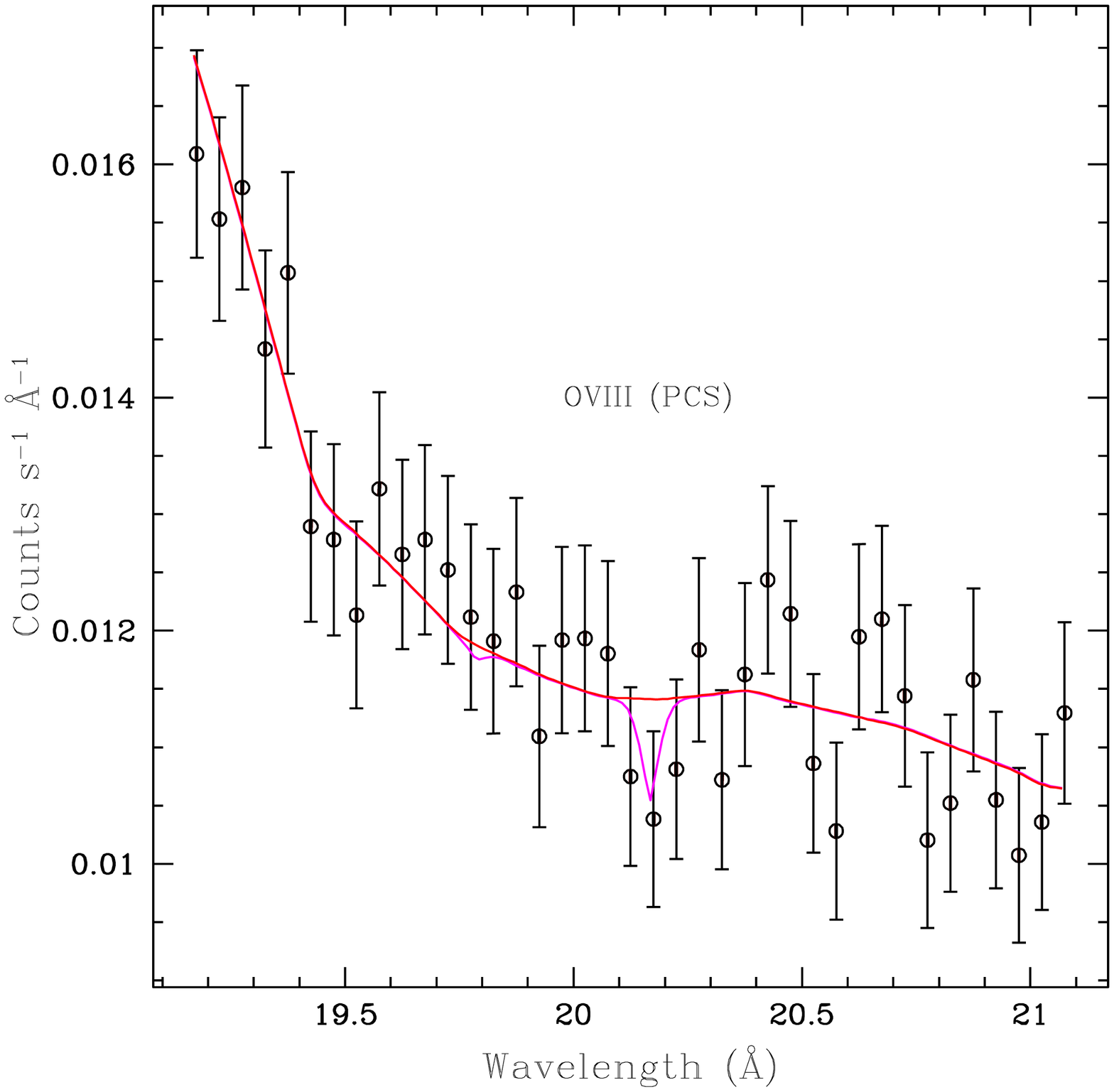}
\caption{Portions of the LETG spectrum of H~2356-309 showing the
  strongest absorption lines from the hot component of the PCS.  The
  best-fitting continuum model is shown in solid thick red, while
  absorption lines from the warm WHIM component are shown in solid
  thin magenta. 
The OVIII K$\alpha$ absorption line at a redshifted wavelength of $\lambda=20.17$ \rm\AA\ is the strongest line of the best-fitting hot-component of the PCS (right panel). 
The other predicted lines are the NeIX K$\alpha$ absorption line at $\lambda=14.29$ \rm\AA\ and two strong L shell transitions from FeXVIII ($\lambda=15.09\rm\AA$, $\lambda=14.20\rm\AA$ rest frame) and FeXVII ($\lambda=15.96\rm\AA$, $\lambda=15.02\AA$ rest frame; left panel).}.
\label{spectraPHASEhotPCS}
  \end{center}
\end{figure*}

For this structure, we found the possible co-existence of two distinct
WHIM phases (Figure ~\ref{contoursPCS}): a warm phase, traced by CV
absorption (Figure~\ref{spectraPHASEwarmPCS}), with $\rm \log{T}
=5.35_{-0.13}^{+0.07} ~K$ and $\rm \log{N_H} = (19.1 \pm 0.2) \times
(Z/Z_{\odot})^{-1} ~cm^{-2}$, and a much hotter and less constrained phase, traced by OVIII absorption  (Figure~\ref{spectraPHASEhotPCS}), with $\rm \log{T} = 6.9^{+0.1}_{-0.8} ~K$ and $\rm \log{N_H} = 20.1^{+0.3}_{-1.7} \times (Z/Z_{\odot})^{-1} ~cm^{-2}$. 

The redshift interval traced by the distribution of galaxies of the PCS, along the line of sight to H~2356-309, is 
$\rm 0.060<z<0.063$. The two WHIM phases tentatively identified here have best-fitting redshifts consistent with each-other, 
and with the PCS redshift interval, namely: $z_{warm} = 0.0623 \pm 0.0005$ and $z_{hot} = 0.063 \pm 0.001$. 

We note that the physical parameters of the putative warm component of the PCS are constrained much better than those of the hot component (Figure~\ref{contoursPCS}). This is because the opacity of absorbing gas to light metal transitions decreases with increasing temperatures. At the best-fitting temperature of the hot component of the PCS, only residual opacity from highly ionized O, Ne and Fe is present (Figure~\ref{spectraPHASEhotPCS}), and the two free model parameters, temperature and N$_H$, becomes highly  correlated (Figure~\ref{contoursPCS}, red, dotted lines), and therefore only poorly constrained. 
At the temperatures of the warm PCS phase, instead, several strong transitions from a number of abundant ions can still produce enough opacity in the data, which makes T and N$_H$ virtually uncorrelated (Figure~\ref{contoursPCS}, cyan solid lines
\footnote{We caution that the opening of the contours at low temperatures is mostly likely a numerical artifact, due to the hitting of the low-temperature end of the model grid, during the error-estimate procedure.} 
). This can be also seen in Figure~\ref{spectraPHASEwarmPCS}, where we
show two portions of the LETG spectrum of H~2356-309 with the strongest absorption lines from the warm component of the PCS, superimposed to the best-fitting continuum (red line) plus warm WHIM model component (blue line). The right panel of Fig.~\ref{spectraPHASEwarmPCS} shows the strongest line of the best-fitting warm-component of the PCS, the CV K$\alpha$ at a redshifted wavelength of $\lambda=42.776$ \AA. However, several other, moderately strong, lines are predicted by this model at shorter wavelengths, and are shown in the left panel of Fig.~\ref{spectraPHASEwarmPCS}, namely the outer shell OVII K$\alpha$ absorption line at $\lambda=22.95\rm\AA$ (superimposed on the OI instrumental edge), and the two inner shell K$\alpha$ transitions from OV at $\lambda=23.74\rm\AA$  ($\lambda=22.35\rm\AA$ rest frame) and OIV at $\lambda=24.17\rm\AA$ ($\lambda=22.75\rm\AA$ rest frame).  The data are consistent with the presence of these lines, which tightly constrain the temperature and column density of the warm component of the PCS. 

\subsubsection{The FSW}
\begin{figure}[!t]
  \begin{center}
\includegraphics[width=0.49\textwidth, angle=0]{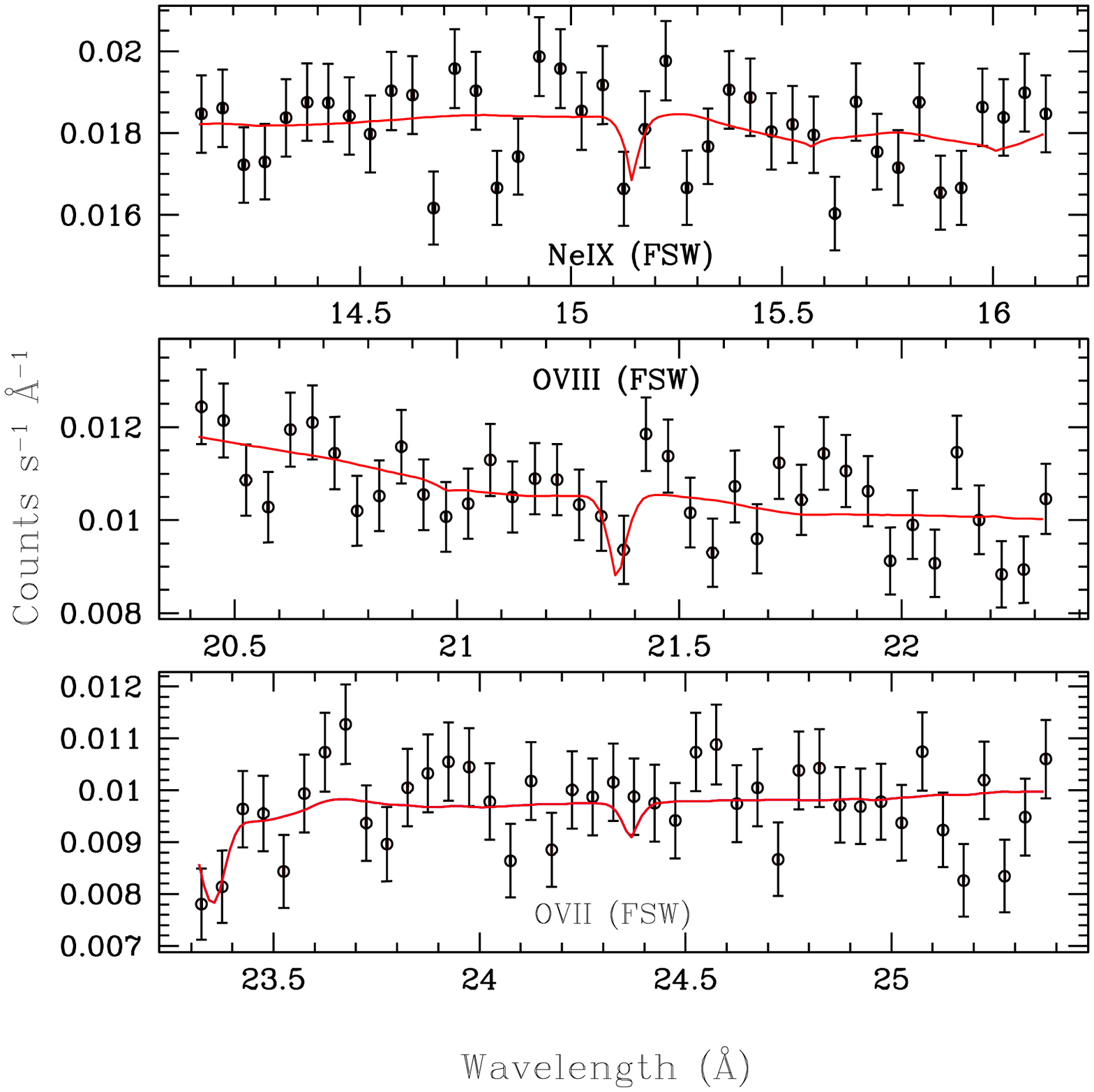}
\caption{Portions of the LETG spectrum of H~2356-309 where the strongest absorption lines from the putative WHIM component of the FSW lie. The OVIII K$\alpha$ absorption line at a redshifted wavelength of $\lambda=21.36$ \AA\ is the strongest line of the best-fitting FSW WHIM model (central panel). 
The other moderately strong predicted line is the NeIX K$\alpha$ absorption line at $\lambda=15.14$ \AA\ (top panel).
In the third panel we also show the region where the OVII K$\alpha$ absorption line of the best-fitting FSW WHIM model falls: the data are clearly incompatible with the presence of such a line.}
\label{spectraPHASE_FSW}
  \end{center}
\end{figure}

For the FSW, the LETG data are consistent with the presence of one hot WHIM component (Figure ~\ref{contoursPCS}), traced mainly by K$\alpha$ OVIII and NeIX absorption (Figure~\ref{spectraPHASE_FSW}). 
The best-fitting WHIM parameter of this component are $\rm \log{T} = 6.6_{-0.2}^{+0.1} ~K$ and $\rm \log{N_H} = 19.8_{-0.8}^{+0.4} ~cm^{-2}$, and, as for the hot component of the PCS, they are only poorly constrained, due to the high best-fitting temperature of the gas. The tighter lower limit on the temperature, compared to that of the hot phase of the PCS, is due to the inconsistency of the data with any OVII K$\alpha$ 
absorption stronger than EW$\ge5.8$ m\AA\ (3$\sigma$ limit; Fig.~\ref{spectraPHASE_FSW}, third panel). 
\footnote{At the redshift of the putative FSW WHIM component, the OVII K$\alpha$ falls at $\lambda=24.25$ \AA, a region of the detector free of instrumental features, unlike the PCS case, where the OVII K$\alpha$ falls at the exact wavelength of the instrumental OI edge.}
. 

The redshift interval traced by the distribution of galaxies of the PCS, along the line of sight to H~2356-309, is 
$\rm 0.127<z<0.129$. The WHIM phase tentatively identified here has
best-fitting redshifts consistent at $1\sigma$ with the low redshift end of the FSW interval, namely: $z_{FSW-WHIM} = 0.126 \pm 0.001$.  

\begin{figure}[!t]
  \begin{center}
\includegraphics[width=0.45\textwidth, angle=0]{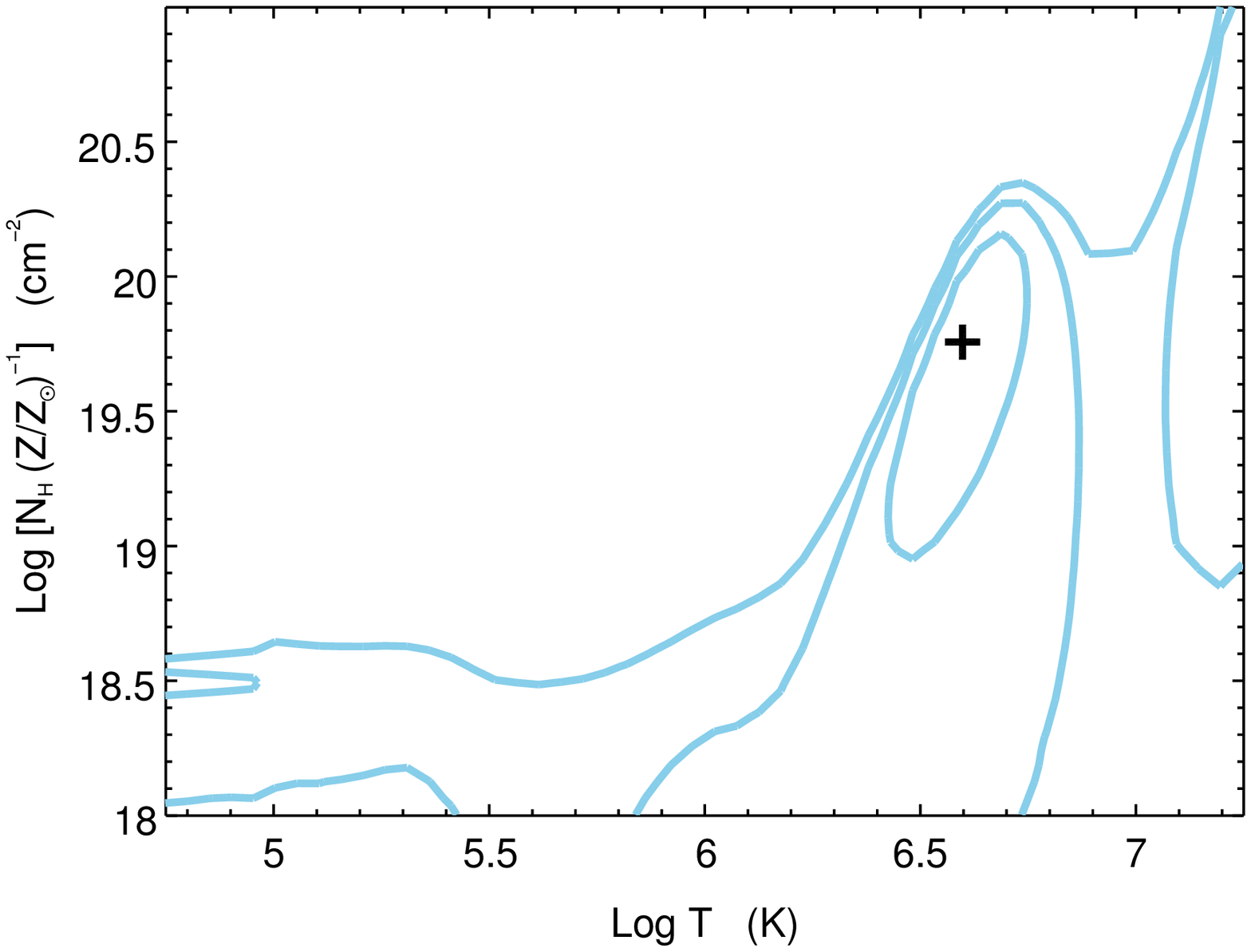}
\caption{68\%, 90\% and 95\%  temperature and equivalent H column density confidence contours for the putative WHIM gas permeating the FSW.}
\label{contoursFSW}
  \end{center}
\end{figure}

\section{Discussion}\label{discussion}
The line of sight to the blazar H~2356-309 is extremely rich in galaxy large-scale filamentary structures (Figure~\ref{LSS}). 
Other than the SW at least two distinct filaments of galaxies cross this line of sight, at average redshifts 
$<z_1>=0.0615$ (the PCS structure) and $<z_2>=0.128$ (the FSW structure). 
These very rich structures have line of sight extensions of $D_1=12.4$ Mpc and $D_2=8$ Mpc, respectively, 
implying Hubble flow velocity ranges of  $\Delta v^{H}_{1} = 890$ km s$^{-1}$ and $\Delta v^{H}_{2} = 580$ km s$^{-1}$. 

The thermal broadening of C or O lines in gas with T$<10^7$ K is $b<85$ km s$^{-1}$ (C) or $b<74$ km s$^{-1}$ (O), negligible 
compared to turbulence induced by peculiar motion of the structures. In simulations \citep[e.g.][]{cen06} WHIM intrinsic 
(i.e. excluding Hubble flow broadening) turbulence are observed to be of $\sim 100$ km/s. 
If  the PCS and the FSW galaxy structures were homogeneously embedded by WHIM, therefore, the Hubble-flow braodening would 
be by far the dominant broadening mechanism, and the gas would imprint metal absorption lines with FWHM$\sim 890$ km 
s$^{-1}$ and FWHM$\sim 580$ km s$^{-1}$, for the PCS and the FSW respectively, easily resolved by the HRC-S/LETG (FWHM=750 
km s$^{-1}$ at 20 \AA\ and FWHM=375 km s$^{-1}$ at 40 \AA). 

On the contrary, unresolved O absoprtion lines in the HRC-LETG must imply that the gas is homogeneously spread over a limited 
portion of these two galaxy structures, extending not more than 4 Mpc along the line of sight (corresponding to Hubble-flow 
broadening of $0.021\AA$ at 20 \AA, i.e. $1\sigma$ of the LETG Line Spread Function)  

\subsection{On the OVII Bearing WHIM or Galaxy Halo Gas}
Our total {\em Chandra} LETG spectrum of H~2356-309 is sensitive to absorption line EW$\ge 14$ m\AA\ at 22 \AA, at 
$\ge 3\sigma$ confidence level. For unsaturated lines ($b\gs 200$ km s$^{-1}$) this corresponds to OVII column densities 
N$_{OVII} \ge 3.4 \times 10^{15} (1+z)^{-1}$ cm$^{-2}$. 
At overdensities $\delta\simeq 50 (1+z)^{-3}$ and temperature T$\simeq 10^6$ K, typical of the bulk ($\sim 50$\%) of
the WHIM density-temperature distribution \citep[e.g. ][]{cen06}, and assuming homogeneity, these columns correspond to line of 
sight extensions of the filament of $D \ge N_b / n_b = [[N_{OVII}/(A_O f_{OVII})] / n_b] (Z/Z_{\odot})^{-1} \simeq 1.4 (Z/Z_{0.1 \odot})^{-1}$ 
Mpc (where we assumed $\rm A_O=8.5\times10^{-4}$, \citealt{grsa}, and a relative fraction of OVII $f_{OVII}=0.9$). 
The extensions of the  PCS and FSW galaxy structures, if entirely permeated by WHIM gas representative of the bulk of its 
density-temperature distribution, should then guarantee the high-significance detection of strong OVII absorbers. 
This has proven to be true only for the SW \citep{fang10} which is the most extended structure along the line of sight with
$D=16.6$ Mpc.

We did not detect strong (i.e. N$_{OVII} \ge  3.4 \times 10^{15} (1+z)^{-1}$ cm$^{-2}$ at $\ge 3\sigma$), unresolved (i.e. $b\le 285$
km s$^{-1}$ at 1$\sigma$) OVII absorption along either of the other two LSSs crossing the line of sight to H~2356-309, the PCS and 
the FSW. 
These non-detections tell us that any ionized gas at typical WHIM temperatures (T$\sim 10^6$ K) embedded in the 4 Mpc cores (i.e. 
imprinting unresolved O lines: see \S 5) 
of the line of sight extent of the PCS and the FSW structures must have overdensities $\delta  \le 18 (Z/Z_{0.1 \odot})^{-1} (1+z)^{-4}$ 
(where the additional $(1+z)^{-3}$ term comes from the $(1+z)^3$ redshift dependency of $<n_b>$). 
For both the PCS ($\delta \le 14 (Z/Z_{0.1 \odot})^{-1}$ at $\ge 3\sigma$), and the FSW ($\delta \le 11 (Z/Z_{0.1 \odot})^{-1}$ at $\ge 
3\sigma$) these overdensities lie on the lower end of the predicted overdensity interval for the bulk of the WHIM, ranging 
between $\delta \simeq 5-50$. This is the opposite of what intuitively expected (though with a large scatter, e.g. Viel et al., 
2005), that richer galaxy LSSs potentially harbor denser WHIM filaments. 

\noindent
Alternatively the OVII bearing gas in these structures, could spread over the large extent of the galaxy super-structure along the line of
sight, and so produce broader and resolved shallow (hence more difficult to detect) lines in the LETG (see \S 5) , or could have 
metallicity lower than 10\% Solar, in either cases relaxing the above limits on the gas overdensities. 
However, any denser WHIM gas at the typical $T\sim 10^6$ K temperature and with line of sight turbulence velocity lower than a few 
hundreds km s$^{-1}$, should have been detected at high significance in the LETG spectrum of H~2356-309. 

An alternative interpretation of the OVII absorption,  is that it could be associated with the hot extended halo
of a single galaxy with small line of sight impact parameter. In
this scenario, if $r$ is the galaxy spherical halo radius, the
impact parameter $d$ must be smaller than $r$, and the line of sight
can only cross a section of the halo $D=2r sin(\alpha)$, where $0
\le \alpha \le \pi/2$ is the angle between the direction of the
galaxy-line of sight impact parameter $d$ and the halo radius $r$ in
the direction of the interception of the halo external boundary with
the line of sight. By averaging over $0 \le \alpha \le \pi/2$, we
get $<D> = 4r/\pi$. With this assumptions the baryon density of the
galaxy halo is given by $n_b = N_b / D \ls [[N_{OVII}/(A_O
f_{OVII})] / (4d/\pi)] (Z/Z_{\odot})^{-1}$. 

The non-detection of OVII bearing gas up to the N$_{OVII} \ge 3.4
\times 10^{15} (1+z)^{-1}$ cm$^{-2}$ 3$\sigma$ limit, allows us to
estimate stringent upper limits on the densities of putative galaxy
halo gas with $T\sim 10^6$ K intercepting the line of sight to
H~2356-309 at the redshifts of the PCS and the FSW structures. By
using an impact parameters $200\lesssim d \lesssim 300$ kpc \citep[][]{stocke}, we obtain
$n_b \simeq (3-5)\times10^{-5} (Z/Z_{0.1 \odot})^{-1}\,\rm cm^{-3}$ for both the PCS and the FSW which is 
lower than the values derived for our galaxy halo or extended 
Local Group gas \citep[e.g.][]{rasmussen03,williams05}. These values 
would be raised at most by a factor of 2 by accounting for non 
constant gas density profile\footnote{We assumed a $\beta-model$
  \citep{cavaliere} gas density distribution with $\beta$ parameter ranging between 0.6 and 0.9
and assumed core radius ranging from 1~kpc to an unrealistically high
400~kpc value.}.

\subsection{The Ionized Gas Content of the PCS and FSW Galaxy Structures} 
As discussed in the previous section, we do not detect significant amount of OVII-bearing (i.e. typical of the bulk WHIM temperature-density distribution) gas in either the PCS or the FSW galaxy super-structures. 
However, at the redshift of these structure, we do marginally detect a number of (individually low-significance) metal absorption lines, from either low-ionization (OVI, OV, CV) or high-ionization (OVIII, NeIX) ions. 
Such ions populate gas with temperatures in the low- or high-end of the WHIM temperature distribution, containing roughly 27\% and 23\% of the predicted WHIM mass, respectively. 
The absorption lines hinted in the LETG spectrum of H~2356-309 at the redshifts of the PCS and the FSW, are all unresolved, which implies line of sight extensions of the absorber $D\ls 4$ Mpc. 

Despite the low statistical significance of each of these individual absorption lines, we were still able to constrain the main physical parameters of the absorbing gas, namely its equivalent H column density and temperature, by modelling the broad-band LETG data with our hybridly ionized WHIM gas model. We identify two different absorbing WHIM phases at the redshift of the PCS, with $\rm\log T = 5.35_{-0.13}^{+0.07}$, $\rm \log N_H = 19.1 \pm 0.2 (Z/Z_{\odot})^{-1}$, and $\rm \log T = 6.9_{-0.8}^{+0.1}$, $\rm \log N_H = 20.1_{-1.7}^{+0.3} (Z/Z_{\odot})^{-1}$, for the warm and the hot phases respectively. 
For the FSW, instead, only one hot phase is tentatively detected, with $\rm \log T = 6.6_{-0.2}^{+0.1}$ and $\rm \log N_H = 19.8_{-0.8}^{+0.4} (Z/Z_{\odot})^{-1}$. 

We can infer baryon densities lower limits for these systems, by
assuming that the absorbers are embedded in their galaxy
superstructures, and have dense cores extending $< 4$ Mpc along the
line of sight (we note that for the two phases in the PCS, assuming
their are physically separated, the total extent can be close to the
entire line of sight extension of the PCS galaxy filament). By
conservatively assuming the $-1\sigma$ $N_H$ value as $N_b$ for
the two PCS WHIM phases we get $n_b(Warm) = N_b(Warm) / D > 6.4 \times
10^{-7} (Z/Z_{\odot})^{-1}$ cm$^{-3}$ ($\delta > 2.7
(Z/Z_{\odot})^{-1}$) and $n_b(hot) = N_b(hot) / D > 2.0 \times 10^{-7}
(Z/Z_{\odot})^{-1}$ ($\delta > 0.9 (Z/Z_{\odot})^{-1}$), while for the
hot phase of the FSW we obtain $n_b = N_b / D > 8.1 \times 10^{-7}
(Z/Z_{\odot})^{-1}$ ($\delta > 2.8 (Z/Z_{\odot})^{-1}$), all consistent with predicted WHIM overdensities. 

\subsection{Number density of OVII absorbers along the line of sight to H~2356-309} 
From our best-fitting WHIM model temperatures for the three putative WHIM absorbers at the redshifts of the PCS and the FSW, we can infer the relative fraction of OVII in each phase. These are $f_{OVII}^{W-PCS} = 0.15$, $f_{OVII}^{H-PCS} = 9 \times 10^{-4}$ and $f_{OVII}^{FSW} = 0.028$. 
These ion fractions correspond to columns of N$_{OVII}^{W-PCS} = 1.6^{+1.8}_{-0.6} \times 10^{15}$ cm$^{-2}$, N$_{OVII}^{H-PCS} = 9.6^{+9.5}_{-9.4} \times 10^{13}$ cm$^{-2}$ and N$_{OVII}^{FSW} = 1.5^{+2.3}_{-1.3} \times 10^{15}$ cm$^{-2}$, or to unsaturated OVII EWs of $7^{+8}_{-3}$ m\AA, $0.42_{-0.41}^{+0.42}$ m\AA\ and $7^{+11}_{-6}$ m\AA, respectively. 

Assuming the best-fitting temperature and N$_H$ values as face values, the lowest OVII equivalent width that we are indirectly probing, through our hybrid-ionization WHIM code, along the line of sight to H~2356-309. is therefore EW(OVII){$\gs 0.4$ m\AA, corresponding to the hot phase of the PCS. 
Combining our three tentative WHIM detections, with the WHIM detection
associated with the SW, leaves us with 4 distinct WHIM systems with
EW(OVII)$\ge 0.4$ m\AA, along a $\Delta z = 0.165$ path length, or a
number density of EW(OVII)$\ge 0.4$ m\AA\ filaments of $dN(EW>0.4)/dz
= 24.2^{+19.2}_{11.6}$ \citep[allowing for the large 1$\sigma$ uncertainties
due to the small number statistics, i.e. ][]{gehrels}. This is
fully consistent with hydrodynamical simulation predictions
\citep[e.g. ][]{cenfang}. 

We can, in principle, also compute the cumulative number density of
OVII filaments per unit redshifts, with EW larger than a given
threshold, for two additional EW thresholds: $\ge 7$ m\AA\ (3
absorbers) and $\ge 25.8$ mAA\ \citep[1 absorber: ][]{fang10}. We get $dN(EW>7)/dz = 18.2^{+17.7}_{9.9}$ and $dN(EW>25.8)/dz = 6.1^{+13.9}_{-5.1}$. 
We plot such derived number density per unit redshift, in Figure~\ref{dndz}, superimposed to the theoretical cumulative $dN(>EW_{thresh})/dz$ curve from \citet{cenfang}. The data-point at 25.8 m\AA, exceeds the predictions by more than 2$\sigma$. 
This could be due to a line of sight selection bias, or may reflect the possibility that one or more of the detected absorption do not actually arise in tenuous WHIM filaments, but in much denser galaxy halos of one (or more) components of the LSSs with line of sight impact parameter of $\sim 200-300$kpc. 
\begin{figure}[!t]
  \begin{center}
\includegraphics[width=0.45\textwidth, angle=0]{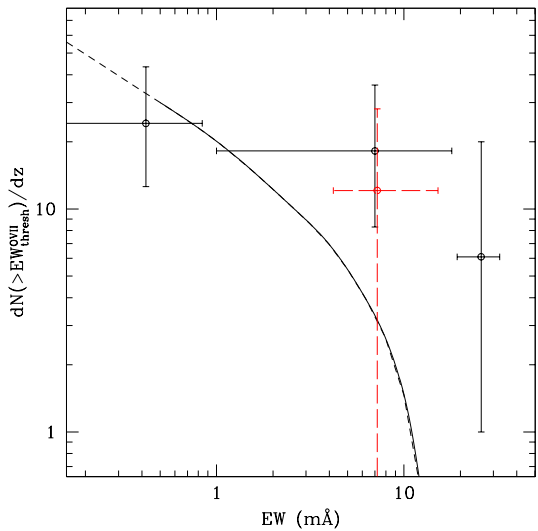}
\caption{Cumulative number density of OVII absorbers per unit
  redshifts compared to the predictions by \citet{cenfang}. The dashed
  line is an extrapolation of the theoretical curve. The dashed data
  point is obtained by excluding the two most marginally detected
  systems: the hot-PCS and FSW.}
\label{dndz}
  \end{center}
\end{figure}
It may finally be that the most marginally detected systems, the Hot-PCS
and the FSW, given their large uncertainties in temperature and column
density may just be explained as occasional statistical
fluctuation. In this case, considering only the SW and Warm-PCS systems,
the number density of EW(OVII)$\ge 7$ m\AA\ filaments would be $dN(EW>7)/dz
= 12.1^{+16.0}_{-11.7}$ which within the large uncertainties is
consistent with the predictions (see Figure~\ref{dndz} dashed data
point).

\subsection{Cosmological Mass Density of the WHIM from the H~2356-309 Line of Sight Absorbers} 
Finally, we derive the Cosmological mass density of the WHIM as measured along the line of sight to H~2356-309, by assuming that all 4 different phases seen at the redshifts of the three SW, PCS and FSW galaxy superstructures are intervening WHIM filaments. 
By propagating in quadrature the large errors associated with the equivalent H column density measurements of the absorbers, and those intrinsic with cosmic variance low-number statistics, we obtain: $\Omega_b^{WHIM} = (0.021^{+0.031}_{-0.018}) (Z/Z_{\odot})^{-1}$. 
This is consistent with the cosmological mass density of intergalactic
Missing Baryons ($\Omega_{b}^{missing-IGM} \simeq 0.016\pm0.005$,
e.g. \citealt{fukugita}) in the local Universe, but assumes solar
metallicities. Metallicities of $\ls 10$\% Solar, would give an
$\Omega_b^{WHIM} $ in excess by $\ge 1\sigma$ compared to
$\Omega_{b}^{missing-IGM}$, again probably indicating an intrinsic
line of sight selection bias, or simply reflecting the possibility
that one or more of the detected absorptions are not imprinted by
tenuous WHIM filaments.  
Again as in previous section we excluded from the analysis the most
uncertain systems Hot-PCS and FSW and estimated an $\Omega_b^{WHIM} =
(0.0026^{+0.0058}_{-0.0018}) (Z/Z_{\odot})^{-1}$ which is consistent
with the predicted missing baryons by assuming $10\%$ solar metallicity.

\section{Conclusions}\label{conclusions}
We have analyzed a long integration (600 ksec) Chandra HRC-S/LETG spectrum
of the blazar H~2356-309 to search and characterize the WHIM in two
large--scale 
galaxy structures along this line of sight and, more specifically,
the ``Pisces-Cetus Supercluster'' (PCS) at z=0.062 and the
``Farther Sculptor Wall'' (FSW) at z=0.128. These are structures more
prominent and more distant with respect to the ``Sculptor Wall'', at z=0.03,
where WHIM was identified in previous works through the detection of the OVII
absorption line.

Although we do not detect significant individual absorption lines in
the PCS nor in the FSW,
the joint analysis of the marginally detected lines (as well as of
stringent upper limits) through a self-consistent hybrid ionization
spectral model, allow us to constrain the physics of the WHIM in the
two farther superstructures. The main results are summarized in the following:

\begin{itemize}

\item
At the redshift of the PCS we identify two distinct phases: a warm phase,
with $\rm \log{T}= 5.35^{+0.07}_{−0.13}~K$
and $\rm log{N_H} = (19.1 \pm 0.2)~cm^{−2}$,
and a much hotter less significant phase, with $\rm log{T} = 6.9^{+0.1}_{−0.8}~ K$
and $\rm log{N_H} = 20.1^{+0.3}_{-1.7}~cm^{-2}$ (1$\sigma$ errors).

\item At the redshift of the FSW only one hot component is hinted
in the data, with $\rm log{T} = 6.6^{+0.1}_{−0.2}~ K$ and
$\rm log{N_H}= 19.8^{+0.4}_{-0.8}~ cm^{-2}$.

\item Under the assumption that the absorbers are embedded in
their galaxy superstructures having baryonic column densities $N_b$
equal to the $-1\sigma$ $N_H$ value, and have dense cores extending
$<$4~Mpc along the line of sight, we can infer conservative lower limits
on the baryons densities in the two systems for the three absorbers. More specifically,
for the two PCS WHIM phases we get $\delta > 2.7 (Z/Z_{\odot})^{-1}$
and $\delta > 0.9 (Z/Z_{\odot})^{-1}$, while for the hot FSW phase we
obtain $\delta > 2.8 (Z/Z_{\odot})^{-1}$, all consistent with predicted WHIM overdensities.

\item By combining the constraints on the OVII absorbers in the PCS
and in the FSW, with the previous detection in the SW \citep{fang10},
we derive the cumulative number density of OVII absorbers per unit
redshift, as a function of the EW(OVII). While at low equivalent
widths (EW(OVII)$>$0.4~m\AA) the absorbers number density is fully
consistent with the predictions of hydrodynamical simulations,
at EW(OVII)$>$10~m\AA \ the inferred absorbers number density
exceeds significantly the theoretical predictions. The latter finding
may result from a line of sight selection bias, or may reflect
the possibility that one or more of the detected absorptions do
not arise in WHIM filaments but in galaxy halos. We considered also
the possibility that the two most marginally detected systems (Hot-PCS
and FSW) may be just statistical fluctuations and not absorbers. In
this case the number density of OVII absorbers per unit redshift would 
agree with the predictions within the large uncertainties.

\item Finally, by combining the measurements in all absorbers
we derive a cosmological mass density of the WHIM of
$\rm \Omega _b^{WHIM}=(0.021^{+0.031}_{-0.018})(Z/Z_{\odot})^{-1}$,
consistent with the cosmological mass density of intergalactic
missing baryons in the local universe. Yet, if the WHIM metallicities
are $<$10\% Solar, then the resulting $\rm \Omega _b^{WHIM}$ is significantly
in excess of the missing baryon density, again possibly indicating a line of
sight selection bias, or reflecting the possibility that one or more
absorbers are not associated with WHIM. The exclusion of the two most
marginally detected systems may bring in agreement the $\rm \Omega
_b^{WHIM}$ value if we assume $0.1\rm Z_\odot$.

\end{itemize}

\begin{acknowledgments}
We thank the anonymous referee for comments and suggestions which
improved the paper presentation. 
We thank Jeremy Drake for useful discussion and support on Chandra LETG calibration. We also thank Yair Krongold for providing the latest version of his hybrid-ionization WHIM models, and Doug Burke for writing the interface routines between Sherpa and our hybrid-ionization models.
FN and LZ acknowledge support from the LTSA grant NNG04GD49G and the ASI-AAE grant I/088/06/0. FN acknoleges support from the FP7-REGPOT-2007-1 EU grant No. 206469.
This research has made use of the NASA/IPAC Extragalactic Database (NED) which is operated by the Jet Propulsion Laboratory, California Institute of Technology, under contract with the National Aeronautics and Space Administration.
\end{acknowledgments}

\bibliography{whim.bib}

\begin{thebibliography}{37}
\expandafter\ifx\csname natexlab\endcsname\relax\def\natexlab#1{#1}\fi

\bibitem[{{Buote} {et~al.}(2009){Buote}, {Zappacosta}, {Fang}, {Humphrey},
  {Gastaldello}, \& {Tagliaferri}}]{buote09}
{Buote}, D.~A., {Zappacosta}, L., {Fang}, T., {Humphrey}, P.~J., {Gastaldello},
  F., \& {Tagliaferri}, G. 2009, \apj, 695, 1351

\bibitem[{{Burns} \& {Batuski}(1984)}]{burns}
{Burns}, J.~O. \& {Batuski}, D.~J. 1984, in ASSL Vol. 111: Clusters and Groups
  of Galaxies, 43--+

\bibitem[{{Cavaliere} \& {Fusco-Femiano}(1976)}]{cavaliere}
{Cavaliere}, A. \& {Fusco-Femiano}, R. 1976, \aap, 49, 137

\bibitem[{{Cen} \& {Fang}(2006)}]{cenfang}
{Cen}, R. \& {Fang}, T. 2006, \apj, 650, 573

\bibitem[{{Cen} {et~al.}(1995){Cen}, {Kang}, {Ostriker}, \& {Ryu}}]{cen95}
{Cen}, R., {Kang}, H., {Ostriker}, J.~P., \& {Ryu}, D. 1995, \apj, 451, 436

\bibitem[{{Cen} \& {Ostriker}(1999)}]{cen}
{Cen}, R. \& {Ostriker}, J.~P. 1999, ApJ, 514, 1

\bibitem[{{Cen} \& {Ostriker}(2006)}]{cen06}
---. 2006, \apj, 650, 560

\bibitem[{{Colless} {et~al.}(2003){Colless}, {Peterson}, {Jackson}, {Peacock},
  {et~al.}}]{colless}
{Colless}, M., {Peterson}, B.~A., {Jackson}, C., {Peacock}, J.~A., {et~al.}
  2003, ArXiv Astrophysics e-prints

\bibitem[{{da Costa} {et~al.}(1988){da Costa}, {Pellegrini}, {Sargent},
  {Tonry}, {Davis}, {Meiksin}, {Latham}, {Menzies}, \& {Coulson}}]{daCosta}
{da Costa}, L.~N., {Pellegrini}, P.~S., {Sargent}, W.~L.~W., {Tonry}, J.,
  {Davis}, M., {Meiksin}, A., {Latham}, D.~W., {Menzies}, J.~W., \& {Coulson},
  I.~A. 1988, ApJ, 327, 544

\bibitem[{{Dav{\' e}} {et~al.}(2001){Dav{\' e}}, {Cen}, {Ostriker}, {Bryan},
  {Hernquist}, {et~al.}}]{dave}
{Dav{\' e}}, R., {Cen}, R., {Ostriker}, J.~P., {Bryan}, G.~L., {Hernquist}, L.,
  {et~al.} 2001, ApJ, 552, 473

\bibitem[{{Fang} {et~al.}(2009){Fang}, {Buote}, {Humphrey}, {Canizares},
  {Zappacosta}, {Maiolino}, {Tagliaferri}, \& {Gastaldello}}]{fang10}
{Fang}, T., {Buote}, D.~A., {Humphrey}, P.~J., {Canizares}, C.~R.,
  {Zappacosta}, L., {Maiolino}, R., {Tagliaferri}, G., \& {Gastaldello}, F.
  2009, ArXiv e-prints

\bibitem[{{Fukugita} {et~al.}(1998){Fukugita}, {Hogan}, \&
  {Peebles}}]{fukugita98}
{Fukugita}, M., {Hogan}, C.~J., \& {Peebles}, P.~J.~E. 1998, \apj, 503, 518

\bibitem[{{Fukugita} \& {Peebles}(2004)}]{fukugita}
{Fukugita}, M. \& {Peebles}, P.~J.~E. 2004, \apj, 616, 643

\bibitem[{{Gehrels}(1986)}]{gehrels}
{Gehrels}, N. 1986, \apj, 303, 336

\bibitem[{{Grevesse} \& {Anders}(1989)}]{grsa}
{Grevesse}, N. \& {Anders}, E. 1989, in American Institute of Physics
  Conference Series, Vol. 183, Cosmic Abundances of Matter, ed.
  {C.~J.~Waddington}, 1--8

\bibitem[{{Hellsten} {et~al.}(1998)}]{hellsten}
{Hellsten}, U. {et~al.} 1998, ApJ, 509, 56

\bibitem[{{Jones} {et~al.}(2004){Jones}, {Saunders}, {Colless}, {Read},
  {Parker}, {et~al.}}]{6dF}
{Jones}, D.~H., {Saunders}, W., {Colless}, M., {Read}, M.~A., {Parker}, Q.~A.,
  {et~al.} 2004, \mnras, 355, 747

\bibitem[{{Kaastra} {et~al.}(2003){Kaastra}, {Lieu}, {Tamura}, {Paerels}, \&
  {den Herder}}]{kaastra}
{Kaastra}, J.~S., {Lieu}, R., {Tamura}, T., {Paerels}, F.~B.~S., \& {den
  Herder}, J.~W. 2003, A\&A, 397, 445

\bibitem[{{Kalberla} {et~al.}(2005){Kalberla}, {Burton}, {Hartmann}, {Arnal},
  {Bajaja}, {Morras}, \& {P{\"o}ppel}}]{karlberla}
{Kalberla}, P.~M.~W., {Burton}, W.~B., {Hartmann}, D., {Arnal}, E.~M.,
  {Bajaja}, E., {Morras}, R., \& {P{\"o}ppel}, W.~G.~L. 2005, \aap, 440, 775

\bibitem[{{Nicastro} {et~al.}(2009){Nicastro}, {Krongold}, {Fields},
  {Conciatore}, {Zappacosta}, {Elvis}, {Mathur}, \& {Papadakis}}]{nicastro10}
{Nicastro}, F., {Krongold}, Y., {Fields}, D., {Conciatore}, M.~L.,
  {Zappacosta}, L., {Elvis}, M., {Mathur}, S., \& {Papadakis}, I. 2009, ArXiv
  e-prints

\bibitem[{{Nicastro} {et~al.}(2008){Nicastro}, {Mathur}, \&
  {Elvis}}]{nicastro08}
{Nicastro}, F., {Mathur}, S., \& {Elvis}, M. 2008, Science, 319, 55

\bibitem[{{Nicastro} {et~al.}(2005{\natexlab{a}}){Nicastro}, {Mathur}, {Elvis},
  {Drake}, {Fiore}, {Fang}, {Fruscione}, {Krongold}, {Marshall}, \&
  {Williams}}]{nicastro05apj}
{Nicastro}, F., {Mathur}, S., {Elvis}, M., {Drake}, J., {Fiore}, F., {Fang},
  T., {Fruscione}, A., {Krongold}, Y., {Marshall}, H., \& {Williams}, R.
  2005{\natexlab{a}}, \apj, 629, 700

\bibitem[{{Nicastro} {et~al.}(2005{\natexlab{b}})}]{nicastro05}
{Nicastro}, F. {et~al.} 2005{\natexlab{b}}, Nature, 433, 495

\bibitem[{{Perna} \& {Loeb}(1998)}]{perna}
{Perna}, R. \& {Loeb}, A. 1998, ApJ, 503, L135+

\bibitem[{{Porter} \& {Raychaudhury}(2005)}]{porter}
{Porter}, S.~C. \& {Raychaudhury}, S. 2005, \mnras, 364, 1387

\bibitem[{{Rasmussen} {et~al.}(2003){Rasmussen}, {Kahn}, \&
  {Paerels}}]{rasmussen03}
{Rasmussen}, A., {Kahn}, S.~M., \& {Paerels}, F. 2003, in astro-ph/0301183,
  109--+

\bibitem[{{Rasmussen} {et~al.}(2007)}]{rasmussen07}
{Rasmussen}, A.~P. {et~al.} 2007, ApJ, 656, 129

\bibitem[{{Scharf} {et~al.}(2000){Scharf}, {Donahue}, {et~al.}}]{scharf}
{Scharf}, C., {Donahue}, M., {et~al.} 2000, ApJ, 528, L73

\bibitem[{{Stocke} {et~al.}(2006){Stocke}, {Penton}, {Danforth}, {Shull},
  {Tumlinson}, \& {McLin}}]{stocke}
{Stocke}, J.~T., {Penton}, S.~V., {Danforth}, C.~W., {Shull}, J.~M.,
  {Tumlinson}, J., \& {McLin}, K.~M. 2006, \apj, 641, 217

\bibitem[{{Tully} {et~al.}(1992){Tully}, {Scaramella}, {Vettolani}, \&
  {Zamorani}}]{tully}
{Tully}, R.~B., {Scaramella}, R., {Vettolani}, G., \& {Zamorani}, G. 1992, ApJ,
  388, 9

\bibitem[{{Viel} {et~al.}(2005){Viel}, {Branchini}, {Cen}, {Ostriker},
  {Matarrese}, {Mazzotta}, \& {Tully}}]{viel}
{Viel}, M., {Branchini}, E., {Cen}, R., {Ostriker}, J.~P., {Matarrese}, S.,
  {Mazzotta}, P., \& {Tully}, B. 2005, \mnras, 360, 1110

\bibitem[{{Werner} {et~al.}(2008){Werner}, {Finoguenov}, {Kaastra},
  {Simionescu}, {Dietrich}, {Vink}, \& {B{\"o}hringer}}]{werner}
{Werner}, N., {Finoguenov}, A., {Kaastra}, J.~S., {Simionescu}, A., {Dietrich},
  J.~P., {Vink}, J., \& {B{\"o}hringer}, H. 2008, \aap, 482, L29

\bibitem[{{Williams} {et~al.}(2005){Williams}, {Mathur}, {Nicastro}, {Elvis},
  {Drake}, {Fang}, {Fiore}, {Krongold}, {Wang}, \& {Yao}}]{williams05}
{Williams}, R.~J., {Mathur}, S., {Nicastro}, F., {Elvis}, M., {Drake}, J.~J.,
  {Fang}, T., {Fiore}, F., {Krongold}, Y., {Wang}, Q.~D., \& {Yao}, Y. 2005,
  \apj, 631, 856

\bibitem[{{Zappacosta}(2005)}]{zapproposal}
{Zappacosta}, L. 2005, in Chandra Proposal, 2041--+

\bibitem[{{Zappacosta}(2006)}]{zapproposal2}
{Zappacosta}, L. 2006, in XMM-Newton Proposal ID \#05043713, 165--+

\bibitem[{{Zappacosta} {et~al.}(2002)}]{zappacosta}
{Zappacosta}, L. {et~al.} 2002, A\&A, 394, 7

\bibitem[{{Zappacosta} {et~al.}(2005)}]{zappacosta2}
---. 2005, MNRAS, 357, 929

\end{thebibliography}
\end{document}